\documentclass[12pt,preprint]{aastex}
\usepackage{url}

\begin{document}

\title{Statistical Analysis of Torus and Kink Instabilities in Solar Eruptions}

\author{Ju Jing\altaffilmark{1}, Chang Liu\altaffilmark{1}, Jeongwoo Lee\altaffilmark{1,2}, Hantao Ji\altaffilmark{3}, Nian Liu\altaffilmark{1}, Yan Xu\altaffilmark{1}, and Haimin Wang\altaffilmark{1}}

\altaffiltext{1}{Center for Solar-Terrestrial Research, New Jersey Institute of Technology, Newark, NJ 07102-1982, USA; ju.jing@njit.edu}
\altaffiltext{2}{School of Space Research, Kyung Hee University, Yongin 17104, Korea}
\altaffiltext{3}{Princeton Plasma Physics Laboratory, Princeton, NJ 08543, USA}

\begin{abstract}
A recent laboratory experiment of ideal magnetohydrodynamic (MHD) instabilities reveals four distinct eruption regimes readily distinguished by the torus instability (TI) and helical kink instability (KI) parameters \citep{Myers2015}. To establish its observational counterpart, we collect 38 solar flares (stronger than GOES class M5 in general) that took place within 45$^{\circ}$ of disk center during 2011$-$2017, 26 of which are associated with a halo or partial halo coronal mass ejection (CME) (i.e., ejective events), while the others are CMEless (i.e., confined events). This is a complete sample of solar events satisfying our selection criteria detailed in the paper. For each event, we calculate decay index $n$ of the potential strapping field above the magnetic flux rope (MFR) in and around the flaring magnetic polarity inversion line (a TI parameter), and the unsigned twist number $T_w$ of the non-linear force-free (NLFF) field lines forming the same MFR (a KI parameter). We then construct a $n-T_w$ diagram to investigate how the eruptiveness depends on these parameters. We find: (1) $T_w$ appears to play little role in discriminating between confined and ejective events; (2) the events with $n\gtrsim0.8$ are all ejective and all confined events have $n\lesssim0.8$. However, $n\gtrsim0.8$ is not a necessary condition for eruption, because some events with $n\lesssim0.8$ also erupted. In addition, we investigate the MFR's geometrical parameters, apex height and distance between footpoints, as a possible factor for the eruptiveness. We briefly discuss the difference of the present result for solar eruptions with that of the laboratory result in terms of the role played by magnetic reconnection.

\end{abstract}


\section{Introduction}

A solar flare is primarily considered to be a low-atmosphere tracer of magnetic explosions/eruptions. In terms of outcome, there are two types of flares: ejective and confined \citep{Moore2001}. Ejective flares are accompanied by coronal mass ejections (CMEs), while confined flares do not have associated CMEs.

A magnetic flux rope (MFR), characterized by a twisted and writhed topological structure, is thought to be a fundamental structure underlying the phenomenon of CMEs. Although the initiation mechanisms are still under debate, it is now common to explain the onset conditions of a MFR eruption in the context of two ideal magnetohydrodynamic (MHD) instabilities, the torus instability (hereafter TI; \citealt{Kliem2006}) and helical kink instability (hereafter KI; \citealt{Torok2004}). TI and KI are mainly controlled by the structural properties of the strapping magnetic field (i.e., the ambient field that runs perpendicular to the MFR) and the guide magnetic field (the ambient field that runs toroidally along the MFR), respectively. Simply put, TI occurs when the strapping field above the MFR declines with height at a sufficiently steep rate, as quantified by decay index $n$ \citep{Torok2005, Kliem2006, Torok2007}. The TI onset criterion of $n\geq n_\mathrm{crit}=1.5$ was first derived analytically by \citet{Bateman1978} and some MHD simulations have found similar values \citep{Kliem2006, Aulanier2010}. A number of other analytical/numerical studies suggest that this critical index $n_\mathrm{crit}$ may lie in a wider range of $0.5<n_\mathrm{crit}<2$ \citep{Fan2007, Demoulin2010, Fan2010, Olmedo2010, Zuccarello2015}. KI, on the other hand, occurs when a MFR is twisted by more than a critical value. The minimum critical twist $\Phi_\mathrm{crit}$ found among analytical/numerical studies is 2.5$\pi$ (corresponding to 1.25 field line windings about the rope axis) \citep[e.g.,][]{Hood1981, Baty2001, Torok2003, Fan2003}. The slow decay of strapping field with height may help confine MFRs, and, in some simulation cases, allows MFRs to build up twist for developing KI \citep{Fan2007}.

Observationally, investigations on what magnetic factors determine the likelihood of ejective/confined eruptions have largely focused on one or two aspects: the decay index $n$ of the potential strapping field \citep[e.g.,][]{LiuY2008, Guo2010, Liu2010a, Nindos2012, Baumgartner2018}, and/or the non-potentiality of active regions (ARs) such as free magnetic energy, relative magnetic helicity, magnetic twist, etc. \citep[e.g.,][]{Nindos2004, Falconer2006, Falconer2009, Tziotziou2012, Lee2016, Toriumi2017}. It has been found in some well-studied cases that confined flares are often hosted by ARs with stronger strapping field and weaker non-potentiality in comparison to ejective ones \citep{Sun2015, Jing2015}. For ejective events, the CME speed shows a positive correlation with the decay index of hosting ARs \citep{Xu2012b, Cui2018}.

It is worth noting that an unprecedented laboratory experiment designed to study the Sun-like line-tied MFRs reveals four distinct eruption regimes which are readily distinguished by the TI and KI parameters (\citealt{Myers2015}; see their Figure 2). In the four regimes MFRs are either eruptive, stable, failed kink (i.e., torus-stable MFRs that exceed the kink threshold fail to erupt), or failed torus (i.e., kink-stable MFRs that exceed the torus threshold fail to erupt). Such an experimental result on the TI and KI has direct implications for eruptions in the solar corona, and its observational counterpart remains to be established, which is the motivation of this study. In this paper, we present the TI vs. KI parameter diagram, established from a statistical study using solar observations together with the coronal field extrapolation techniques. The goal of this study is to improve our understanding of the requirements for a solar eruption: what the trigger/driver mechanisms might be, and what, if any, onset criteria must be reached.

\section{Sample Selection and Methods}

We examined NOAA GOES soft X-ray (SXR) flare reports to search for major flares (stronger than GOES class M5 in general) that occurred within 45$^{\circ}$ of the disk center over a seven-year period from January 2011 to December 2017. Due to the small sample size of confined flares, we relaxed the SXR class requirement from M5 to M4 for confined flares. To avoid the over-representation of a certain flare-productive AR, at most two flares per AR were included into the samples, the one of the greatest SXR magnitude, and the one nearest to the disk center.

For each flare, its CME association was determined by reference to the \emph{LASCO} CME catalog \citep{Gopalswamy2009a}. We regarded a flare as ejective if the following criteria are fulfilled: (1) the CME onset time at $R_{\odot}$ extrapolated backward from the CME heigh-time profile reasonably agrees with the flare onset time; and (2) the position angle of the CME agrees with the quadrant on the Sun in which the flare occurred. When a flare-CME association is identified, we also refer to the \emph{LASCO} CME catalog for the CME kinetic energy and use it as a CME parameter \citep{Vourlidas2010}. We then excluded those ejective flares from the samples if their associated CMEs are neither halo nor partial-halo, because the other types (for example, a jet-type) of CMEs may not be compatible with a MFR topology. We regarded a flare as confined if there are no CMEs in temporal and spatial proximity as described above.

The sample selection requirements led us to a total of 38 flares (26 ejective and 12 confined) from 27 different ARs. Table 1 summarizes the properties of the flares and the CME kinetic energy $E_\mathrm{CME}$ (if ejective). The last five columns list free magnetic energy $E_\mathrm{free}$ (a measure of the non-potentiality of ARs), apex height $h_\mathrm{apex}$ and footpoint distance $d$ of MFRs, decay index $n$ of the strapping field (a TI parameter; see \S3.1), and twist number $T_w$ of the MFR (a KI parameter; see \S3.2). The calculation of all these parameters involves coronal magnetic field measurements, which are hardly accessible from observations but can be reconstructed using a technique called coronal field extrapolation.

For each event, we used the last available vector magnetograms (\verb#hmi.sharp_cea_720s#) of the AR prior to the flare, obtained by the the Helioseismic and Magnetic Imager (HMI; \citealt{Schou2012}) on board the Solar Dynamics Observatory (SDO), as the boundary conditions for the coronal field extrapolation. The magnetograms were re-mapped using a cylindrical equal area (CEA) projection and presented as ($B_r$,$B_\theta$,$B_\phi$) in heliocentric spherical coordinates corresponding to ($B_z$,−$B_y$,$B_x$) in heliographic coordinates \citep{Sun2013}. We re-binned the magnetograms to 1\arcsec\ pixel intervals and pre-processed the data towards the force-free conditions \citep{Wiegelmann2006}. Then we performed the three-dimensional (3D) nonlinear force-free (NLFF) field and potential field extrapolations with the weighted optimization method \citep{Wiegelmann2004} and the method of \citet{Alissandrakis1981}, respectively. In particular, the weighted optimization method of \citet{Wiegelmann2004} is an implementation of the original optimization algorithm of \citet{Wheatland2000}, and has been optimized for the SDO/HMI magnetogram data \citep{Wiegelmann2012}. The x- and y-dimensions of the 3D computational domain $V$ vary from case to case according to ARs's size to cover not only the major portion of ARs but also the plage regions surrounding the ARs, and the z-dimension of $V$ is set to be 200\arcsec ($\sim$ 145~Mm) in all the cases.

To assess the performance of NLFF field extrapolations, for each event, we computed the $\mathrm{<CW\sin\theta>}$ \footnote {$<\mathrm{CW\sin\theta}>=(\sum_i|\textbf{J}_i|\theta_i)/\sum_i |\textbf{J}_i|$, where $\theta$ is the angle between $\textbf{B}$ and $\textbf{J}$.} metric and the $<|f_i|>$ \footnote{$|f_i|=|(\nabla\cdot \textbf{B})_i|/(6|\textbf{B}|_i/\triangle x)$, where $\triangle$x is the grid
spacing.} metric (see \citealt{DeRosa2009} for details). Briefly, $<\mathrm{CW\sin\theta}>$ and $<|f_i|>$ measure how well the force-free and divergence-free conditions are satisfied in the NLFF field models, with perfectly force-free and divergence-free fields having $<\mathrm{CW\sin\theta}>=0$ and $<|f_i|>=0$, respectively. The average of all values of $<CW\sin \theta>$ over our 38 samples is 0.13$\pm$0.04, and that of $<|f_i|>$ is (4.4$\pm$1.1)$\times$10$^{-4}$, suggesting a well satisfied force-free and divergence-free condition.

Based on both NLFF and potential fields, we estimated free magnetic energy $E_\mathrm{free}$ over the 3D computational domain $V$ as follows:
\begin{equation}
E_\mathrm{free}=E_\mathrm{NLFF}-E_\mathrm{p}=\int\frac{B_\mathrm{NLFF}^2}{8\pi}dV - \int\frac{B_\mathrm{p}^2}{8\pi}dV,
\end{equation}
where the superscripts $\mathrm{NLFF}$ and $\mathrm{p}$ represent the NLFF field and the potential field, respectively. $E_\mathrm{free}$ calculated in this way is regarded as the maximum energy available for powering flares/CMEs in a given AR. The uncertainty of $E_{free}$ caused by the boundary conditions was previously evaluated by the code testing conducted by \citet{Wiegelmann2010}, in which the same extrapolation code was applied to a well known semi-analytic test case of \citet{Low1990}. It is found that, for example, for a relatively high noise level of 5\% in the transverse field, the reconstructed NLFF field is able to reproduce the reference field with an underestimation of 12-14\% in $E_\mathrm{NLFF}$ \citep{Wiegelmann2010}, which may lead to an uncertainty of $\sim$39\% in $E_\mathrm{free}$. The value of $E_\mathrm{free}$ calculated from our NLFF and potential fields with a 39\% uncertainty is listed in Table 1.

The TI and KI parameters were inferred from the NLFF field model and potential field model, respectively, as described in the following section.

\section{TI and KI Parameters}
The estimation of TI and KI parameters is illustrated in Figure 1 and Figure 2, using the ejective M6.5 (SOL2015-06-22T18:23) flare of AR 12371 and the confined M4.2 (SOL2015-03-12T14:08) flare of AR 12297 as examples, respectively.

\subsection{TI Parameter: Decay Index $n$}
TI is triggered by the force imbalance in the vertical direction (the upward ``hoop" force versus the downward strapping force acting on the MFR), and is often quantified by the decay index $n$ of the potential strapping field \citep{Bateman1978, Kliem2006}:
\begin{equation}
n= - \frac{\partial \log(B_{ext})}{\partial \log(h)},
\end{equation}
\noindent where $B_{ext}$ denotes the external strapping field at a geometrical height $h$ above the surface. For each event, we first followed the procedures described by \citet{Sun2015} to identify the flaring polarity inversion line (FPIL) mask which demarcates the AR core field where the MFR resides and the instability initiates. The HMI $B_z$ map together with the UV 1600~\AA\ image taken near the flare peak time by the Atmospheric Imaging Assembly (AIA; \citealt{Lemen2012}) on board SDO was used in this step. Figure 1a (2a) shows the composite image of both $B_z$ and AIA 1600~\AA\ maps of the ejective M6.5 (confined M4.2) flare, superimposed with the yellow contour of the FPIL mask. The horizontal component of the potential field directly above the FPIL is used to approximate the strapping magnetic field. Figure 1c (2c) shows the corresponding profile of $n$ (averaged over all FPIL mask pixels) as a function of height $h$. The mean value of $n$ at the MFR's apex height $h_\mathrm{apex}$ with the $\pm1\sigma$ uncertainty calculated over the FPIL mask is listed in Table 1, .

\subsection{KI Parameter: Twist Number $T_w$}
In our NLFF field models, the twist number $T_w$ \citep{Berger2006} is computed as:
\begin{equation}
T_w=\int_L\frac{\mu_0 \mathbf{J} \cdot \mathbf{B}/ |\mathbf{B}|}{4\pi |\mathbf{B}|}dl=\int_L\frac{\nabla \times \mathbf{B} \cdot \mathbf{B}}{4\pi |\mathbf{B}|^2}dl = \frac{1}{4\pi}\int_L \alpha dl,
\end{equation}
where $\alpha$ is the force-free parameter, and $\frac{\nabla \times \mathbf{B} \cdot \mathbf{B}}{4\pi |\mathbf{B}|^2}$ is thought as a local density of twist along each individual field line. $T_w$ is a good approximation to the traditional physical concept of twist $\Phi$ (i.e., winding of magnetic field lines around an axis) in the vicinity of the axis of a nearly cylindrically symmetric MFR \citep{Liu2016}. While the computation of $\Phi$ requires the correct determination of the axis, $T_w$ can be computed straightforwardly for any field lines without restoring the geometry of a MFR, and thus providing a convenient means to quantify KI in practice. To calculate $T_w$ we used the code developed by \citet{Liu2016}, which is available online \footnote {http://staff.ustc.edu.cn/~rliu/qfactor.html}.

We applied the code to the extrapolated 3D NLFF field to produce the $T_w$ map in which each pixel is assigned by the $T_w$ value of the NLFF field line threading from this pixel. Figure 1b(2b) shows the $T_w$ map of the ejective M6.5 (confined M4.2) flare of AR 12371 (AR 12297). Near the two ends of the FPIL region, we see two conjugate regions with enhanced $T_w$ of the same sign, which are considered to host the footpoints of the MFR. The mean unsigned twist number $|T_w|$ of the NLFF field lines forming the MFR (derived from NLFF field; see the bottom panels of Figures 1 and 2) and its $\pm1\sigma$ uncertainty are listed in Table 1.

\section{Results}
Figure 3 shows the scatter diagram of TI parameters $n$ vs. KI parameters $T_w$ for the 38 flares. The black symbols represent the confined flares and the rest, ejective flares. For ejective flares, the color is assigned according to the associated CME's kinetic energy. At a glance, our result does not clearly show the four distinct eruption regimes found by the laboratory experiment (Figure 2 of Myers et al. 2015). It is partly due to the fact that the confined and ejective flares are not clearly distinguished in terms of $T_w$. Instead, we see the clustering of the confined CMEless flares in the lower part of the diagram (0.2~$\lesssim$~n~$\lesssim$~0.7), while the ejective flares spread out over most of the $n$ range (0.2~$\lesssim$~n~$\lesssim$~1.6). Note that the 12 flares with $n\gtrsim0.8$ are all ejective, in which sense we can regard this as a sufficient condition for CME. However, not all flares of $n\lesssim0.8$ are confined. Only 12 out of the 26 flares with $n\lesssim0.8$ are confined and the rest 14 flares are ejective. Thus the criterion, $n>n_\mathrm{crit}\simeq0.8$ found here is not a necessary condition for CME. Note also that this value of $n_\mathrm{crit}\simeq0.8$ found here is much lower than those typically cited in other solar studies  ($n_\mathrm{crit}\simeq1.1-1.3$ in \citealt{Demoulin2010}, for instance), although it agrees well with the critical decay index found in the laboratory experiment performed by \citet{Myers2015}. We presume that the difference arises from the fact that the decay indices were often evaluated for large loops (typically in the height range of $42−-105$~Mm) in the previous solar studies whereas the MFRs with lower heights are included in the present statistical study.

Based on the experimentally measured TI vs. KI parameter diagram, \citet{Myers2015} report a previously unknown instability regime $--$ failed torus. The ``failed torus'' events occur when the guide magnetic field interacts with electric currents in the MFR to produce a dynamic tension force which brakes the ascension in the torus-unstable region. Our limited samples, however, do not show the presence of this regime. Instead, all the MFRs that exceed a certain torus threshold, $\sim$0.7 in our cases, are developing into CMEs. Presumably the dynamic tension force in solar cases is too weak to halt eruptions.

The top panels of Figure 4 show the histograms of $|T_w|$, $n$, and $E_\mathrm{free}$ for both ejective and confined flares. To investigate the MFR geometry as a possible factor for the eruptiveness, we also compared the histograms of $h_\mathrm{apex}$, distance $d$ between the MFR footpoints, and $h_\mathrm{apex}/d$ for ejective and confined flares in the bottom panels of Figure 4. For a quantitative comparison between the ejective and confined samples, we performed the Student’s t-test to determine the t-statistic ($t$; a ratio of the difference between two groups to the difference within the groups) and its significance ($\alpha$; the probability that the results occurred by random chance) for each of the parameters. Briefly speaking, the larger the t-statistic, the more difference there is between the two groups; The lower the significance, the more confident one can replicate the results. As one might expect, the most appreciable segregation between the two groups is in the histograms of $n$ with $t$=2.337 and $\alpha$=0.025. That is, the null hypothesis (i.e., there is no difference in mean $n$ between the ejective and confined flares) can be rejected at the 100(1-$\alpha$)\%= 97.5\% level of confidence. By comparison, the role of $T_w$ in distinguishing between ejective and confined groups is questionable ($t$=0.995 and $\alpha$=0.32). Based on this result, we conclude that the TI rather than the KI plays a more important role in differentiating between the ejective and the confined flares.

To illustrate the relationship between the MFR geometry and the strapping effect, Figure 5 shows the scatter diagrams of (a) $h_\mathrm{apex}$ vs. $d$, (b) $n$ vs. $d$, (c) $n$ vs. $h_\mathrm{apex}$, and (d) $n$ vs. $h_\mathrm{apex}/d$. The linear Pearson correlation coefficient (CC) and the probability of obtaining a certain CC by chance ($P_{CC}$) are given in the each panel. We see a moderate positive linear correlation between $h_\mathrm{apex}$ and $d$ with a CC of 0.62 (Figure 5a). The linear fit to these two data pairs is $h_\mathrm{apex}=2.65+0.26\times d$, suggesting that MFRs in our solar cases are of a more flat-arched structure or are only a minor segment of a circular structure. A strong positive correlation between $n$ and $h_\mathrm{apex}$ with a CC of 0.76 is shown in Figure 5c. This should not be surprising, as the strapping magnetic field decays with height so that a low-lying/high-lying MFR is usually relevant to a stronger/weaker strapping effect.

\section{Discussion and Conclusion}

The previous laboratory experiment reveals that the eruptiveness of MFRs is dependant on the interplay between the TI and KI, as represented by the $n-T_w$ diagram. In this paper we intended to establish a solar counterpart to the diagram, by which we may be able to tell the likelihood of a CME based on the observed $n$ and $T_w$ parameters. The key results are summarized and discussed as follows:

First, the TI quantified by $n$ appears to play an important role in differentiating between ejective and confined flares. However, the TI onset criteria ($n\geq n_\mathrm{crit}=\sim0.75$) found here is not a necessary condition for CMEs. Some MFRs in the TI-stable regime still manage to break through the strong strapping field and evolve into CMEs. It therefore implies that an additional trigger and driving mechanism may be involved in solar eruptions. A very likely candidate for the alternative process is magnetic reconnection during solar flares. Actually there are a number of analytical/numerical models invoking magnetic reconnection in the mechanism of CMEs. For instance, in the magnetic breakout model \citep{Antiochos1999}, magnetic reconnection leads to the progressive transformation of the magnetic configuration, allowing a MFR to burst open. In the tether-cutting reconnection model \citep{Moore2001}, magnetic reconnection below a MFR ``cut''s the ``tether''s of the strapping field to unleash CMEs. Such non-ideal MHD processes are absent in the laboratory experiment which was designed to simulate eruptions solely in terms of an ideal MHD process.

Second, it is unclear in this study whether the KI represented by $T_w$ is a major factor for solar eruption. Two MFRs with the highest value of $T_w>1.2$ erupted, but many other MFRs with smaller values of $T_w$ were also able to erupt, and we tend to believe that KI is less influential. We consider two possible caveats. The first concerns the ongoing debate whether a helical magnetic structure pre-exists before an eruption \citep{Low1994, Chen1989, Fan2004} or is formed in the course of an eruption via magnetic reconnection \citep{vanBallegooijen1989, Amari2000, MacNeice2004}. There are observational evidence in favor of each scenario \citep[e.g.,][]{Dere1999, Qiu2007, Liu2010b, Song2014, Wang2015, Yan2015, Gopalswamy2017}. In the latter case, it is not surprising that $T_w$ derived from the pre-eruption magnetic field may be underestimated and can not correctly predict the eruptiveness. The second possibility is that helical KI could result in the deformation of a MFR, but may not allow a huge expansion of the MFR to produce a CME \citep{Green2018}. In this sense we may consider that KI might be capable of initiating a filament eruption and a flare, but may not be the key factor in driving a CME into the heliosphere.

Third, the laboratory experiment by \citet{Myers2015} shows that there can be both failed TI and failed KI events. Namely, MFRs have more difficulty in eruption than solar community believed. This is contrary to our results that even the TI-stable ($n<0.75$) ones can erupt and CMEs can occur regardless of the KI parameter $T_w$. As mentioned earlier, we speculate that magnetic reconnection, which was absent in the laboratory experiment, may be the factor causing the differences between the laboratory and the present solar observations, if it alleviates the difficulties in eruption.

The differences between the laboratory results and our results may also arise from multiple sources of assumptions and approximations of this study in contrast to the lab experiment. In the present study, the TI and KI parameters $n$ and $T_w$ are not directly measured in observations, but rather estimated from MFRs in NLFF field models. The identification of MFRs relies on the quality of NLFF field extrapolation. Although the up-to-date NLFF field extrapolation technique employed here was evaluated thoroughly in comparison with a 3D radiative MHD model and was found to offer a reasonably high accuracy of the coronal field reconstruction \citep{Wiegelmann2010a, Wiegelmann2010, Fleishman2017}, the direct validation of NLFF fields still cannot be performed due to the lack of the coronal magnetic field diagnostics. We'd like to add a caution that NLFF field extrapolation has intrinsic limitations associated with the force-free assumption and is subject to numerous uncertainties in the data reduction and modeling process which are not reflected in our results. It may be that $n$ and/or $T_w$ could not be accurately calculated under the observational limits. In addition, the KI parameter $T_w$ is derived from and averaged over individual field lines, assuming that it's related to the winding of field lines around the axis, but actually the twist of a MFR could be underestimated by its built-in assumption.

Finally, we'd like to mention that the present statistical study is a step forward to access the role of the TI and KI in solar eruptions. Detailed studies of the pre-to-post flare magnetic configuration are also needed to better understand the underlying physics, which will be conducted in the future.

\begin{acknowledgements}
We thank the anonymous referee for the constructive comments. We thank the NASA SDO team for HMI and AIA data. HMI and AIA are instruments on board SDO, a mission for NASA's Living with a Star program. We thank CDAW Data Center at Goddard Space Flight Center for providing the CME catalog, which is supported by NASA's Living with a Star program and the SOHO project. We thank Dr. Rui Liu for developing the code to calculate twist number. J.J., N.L., Y.X., C.L. and H.W. were supported by NASA grants NNX16AF72G, 80NSSC17K0016 (Grand Challenge) and 80NSSC18K0673 (GI), NSF grants AGS 1408703, and 1539791. J.L. was supported by the BK21 plus program through the National Research Foundation (NRF) funded by the Ministry of Education of Korea to Kyung Hee University.

\end{acknowledgements}


\begin{thebibliography}{}
\expandafter\ifx\csname natexlab\endcsname\relax\def\natexlab#1{#1}\fi

\bibitem[{{Alissandrakis}(1981)}]{Alissandrakis1981}
{Alissandrakis}, C.~E. 1981, \aap, 100, 197

\bibitem[{{Amari} {et~al.}(2000){Amari}, {Luciani}, {Mikic}, \&
  {Linker}}]{Amari2000}
{Amari}, T., {Luciani}, J.~F., {Mikic}, Z., \& {Linker}, J. 2000, \apjl, 529,
  L49

\bibitem[{{Antiochos} {et~al.}(1999){Antiochos}, {DeVore}, \&
  {Klimchuk}}]{Antiochos1999}
{Antiochos}, S.~K., {DeVore}, C.~R., \& {Klimchuk}, J.~A. 1999, \apj, 510, 485

\bibitem[{{Aulanier} {et~al.}(2010){Aulanier}, {T{\"o}r{\"o}k}, {D{\'e}moulin},
  \& {DeLuca}}]{Aulanier2010}
{Aulanier}, G., {T{\"o}r{\"o}k}, T., {D{\'e}moulin}, P., \& {DeLuca}, E.~E.
  2010, \apj, 708, 314

\bibitem[{{Bateman}(1978)}]{Bateman1978}
{Bateman}, G. 1978, {MHD instabilities}

\bibitem[{{Baty}(2001)}]{Baty2001}
{Baty}, H. 2001, \aap, 367, 321

\bibitem[{{Baumgartner} {et~al.}(2018){Baumgartner}, {Thalmann}, \&
  {Veronig}}]{Baumgartner2018}
{Baumgartner}, C., {Thalmann}, J.~K., \& {Veronig}, A.~M. 2018, \apj, 853, 105

\bibitem[{{Berger} \& {Prior}(2006)}]{Berger2006}
{Berger}, M.~A., \& {Prior}, C. 2006, Journal of Physics A Mathematical
  General, 39, 8321

\bibitem[{{Chen}(1989)}]{Chen1989}
{Chen}, J. 1989, \apj, 338, 453

\bibitem[{Cui {et~al.}(2018)Cui, Wang, Xu, \& Liu}]{Cui2018}
Cui, Y., Wang, H., Xu, Y., \& Liu, S. 2018, Journal of Geophysical Research:
  Space Physics, 123, 1704

\bibitem[{{De Rosa} {et~al.}(2009){De Rosa}, {Schrijver}, {Barnes}, {Leka},
  {Lites}, {Aschwanden}, {Amari}, {Canou}, {McTiernan}, {R{\'e}gnier},
  {Thalmann}, {Valori}, {Wheatland}, {Wiegelmann}, {Cheung}, {Conlon},
  {Fuhrmann}, {Inhester}, \& {Tadesse}}]{DeRosa2009}
{De Rosa}, M.~L., {Schrijver}, C.~J., {Barnes}, G., {et~al.} 2009, \apj, 696,
  1780

\bibitem[{{D{\'e}moulin} \& {Aulanier}(2010)}]{Demoulin2010}
{D{\'e}moulin}, P., \& {Aulanier}, G. 2010, \apj, 718, 1388

\bibitem[{{Dere} {et~al.}(1999){Dere}, {Brueckner}, {Howard}, {Michels}, \&
  {Delaboudiniere}}]{Dere1999}
{Dere}, K.~P., {Brueckner}, G.~E., {Howard}, R.~A., {Michels}, D.~J., \&
  {Delaboudiniere}, J.~P. 1999, \apj, 516, 465

\bibitem[{{Falconer} {et~al.}(2006){Falconer}, {Moore}, \&
  {Gary}}]{Falconer2006}
{Falconer}, D.~A., {Moore}, R.~L., \& {Gary}, G.~A. 2006, \apj, 644, 1258

\bibitem[{{Falconer} {et~al.}(2009){Falconer}, {Moore}, {Gary}, \&
  {Adams}}]{Falconer2009}
{Falconer}, D.~A., {Moore}, R.~L., {Gary}, G.~A., \& {Adams}, M. 2009, \apjl,
  700, L166

\bibitem[{{Fan}(2010)}]{Fan2010}
{Fan}, Y. 2010, \apj, 719, 728

\bibitem[{{Fan} \& {Gibson}(2003)}]{Fan2003}
{Fan}, Y., \& {Gibson}, S.~E. 2003, \apjl, 589, L105

\bibitem[{{Fan} \& {Gibson}(2004)}]{Fan2004}
---. 2004, \apj, 609, 1123

\bibitem[{{Fan} \& {Gibson}(2007)}]{Fan2007}
---. 2007, \apj, 668, 1232

\bibitem[{{Fleishman} {et~al.}(2017){Fleishman}, {Anfinogentov}, {Loukitcheva},
  {Mysh'yakov}, \& {Stupishin}}]{Fleishman2017}
{Fleishman}, G.~D., {Anfinogentov}, S., {Loukitcheva}, M., {Mysh'yakov}, I., \&
  {Stupishin}, A. 2017, \apj, 839, 30

\bibitem[{{Gopalswamy} {et~al.}(2017){Gopalswamy}, {Yashiro}, {Akiyama}, \&
  {Xie}}]{Gopalswamy2017}
{Gopalswamy}, N., {Yashiro}, S., {Akiyama}, S., \& {Xie}, H. 2017, \solphys,
  292, 65

\bibitem[{{Gopalswamy} {et~al.}(2009){Gopalswamy}, {Yashiro}, {Michalek},
  {Stenborg}, {Vourlidas}, {Freeland}, \& {Howard}}]{Gopalswamy2009a}
{Gopalswamy}, N., {Yashiro}, S., {Michalek}, G., {et~al.} 2009, Earth Moon and
  Planets, 104, 295

\bibitem[{{Green} {et~al.}(2018){Green}, {T{\"o}r{\"o}k}, {Vr{\v s}nak},
  {Manchester}, \& {Veronig}}]{Green2018}
{Green}, L.~M., {T{\"o}r{\"o}k}, T., {Vr{\v s}nak}, B., {Manchester}, W., \&
  {Veronig}, A. 2018, \ssr, 214, 46

\bibitem[{{Guo} {et~al.}(2010){Guo}, {Ding}, {Schmieder}, {Li},
  {T{\"o}r{\"o}k}, \& {Wiegelmann}}]{Guo2010}
{Guo}, Y., {Ding}, M.~D., {Schmieder}, B., {et~al.} 2010, \apjl, 725, L38

\bibitem[{{Hood} \& {Priest}(1981)}]{Hood1981}
{Hood}, A.~W., \& {Priest}, E.~R. 1981, Geophysical and Astrophysical Fluid
  Dynamics, 17, 297

\bibitem[{{Jing} {et~al.}(2015){Jing}, {Xu}, {Lee}, {Nitta}, {Liu}, {Park},
  {Wiegelmann}, \& {Wang}}]{Jing2015}
{Jing}, J., {Xu}, Y., {Lee}, J., {et~al.} 2015, Research in Astronomy and
  Astrophysics, 15, 1537

\bibitem[{{Kliem} \& {T{\"o}r{\"o}k}(2006)}]{Kliem2006}
{Kliem}, B., \& {T{\"o}r{\"o}k}, T. 2006, Physical Review Letters, 96, 255002

\bibitem[{{Lee} {et~al.}(2016){Lee}, {Liu}, {Jing}, \& {Chae}}]{Lee2016}
{Lee}, J., {Liu}, C., {Jing}, J., \& {Chae}, J. 2016, \apjl, 831, L18

\bibitem[{{Lemen} {et~al.}(2012){Lemen}, {Title}, {Akin}, {Boerner}, {Chou},
  {Drake}, {Duncan}, {Edwards}, {Friedlaender}, {Heyman}, {Hurlburt}, {Katz},
  {Kushner}, {Levay}, {Lindgren}, {Mathur}, {McFeaters}, {Mitchell}, {Rehse},
  {Schrijver}, {Springer}, {Stern}, {Tarbell}, {Wuelser}, {Wolfson}, {Yanari},
  {Bookbinder}, {Cheimets}, {Caldwell}, {Deluca}, {Gates}, {Golub}, {Park},
  {Podgorski}, {Bush}, {Scherrer}, {Gummin}, {Smith}, {Auker}, {Jerram},
  {Pool}, {Soufli}, {Windt}, {Beardsley}, {Clapp}, {Lang}, \&
  {Waltham}}]{Lemen2012}
{Lemen}, J.~R., {Title}, A.~M., {Akin}, D.~J., {et~al.} 2012, \solphys, 275, 17

\bibitem[{{Liu} {et~al.}(2010{\natexlab{a}}){Liu}, {Lee}, {Jing}, {Liu},
  {Deng}, \& {Wang}}]{Liu2010a}
{Liu}, C., {Lee}, J., {Jing}, J., {et~al.} 2010{\natexlab{a}}, \apjl, 721, L193

\bibitem[{{Liu} {et~al.}(2010{\natexlab{b}}){Liu}, {Liu}, {Wang}, {Deng}, \&
  {Wang}}]{Liu2010b}
{Liu}, R., {Liu}, C., {Wang}, S., {Deng}, N., \& {Wang}, H. 2010{\natexlab{b}},
  \apjl, 725, L84

\bibitem[{{Liu} {et~al.}(2016){Liu}, {Kliem}, {Titov}, {Chen}, {Wang}, {Wang},
  {Liu}, {Xu}, \& {Wiegelmann}}]{Liu2016}
{Liu}, R., {Kliem}, B., {Titov}, V.~S., {et~al.} 2016, \apj, 818, 148

\bibitem[{{Liu}(2008)}]{LiuY2008}
{Liu}, Y. 2008, \apjl, 679, L151

\bibitem[{{Low}(1994)}]{Low1994}
{Low}, B.~C. 1994, Physics of Plasmas, 1, 1684

\bibitem[{{Low} \& {Lou}(1990)}]{Low1990}
{Low}, B.~C., \& {Lou}, Y.~Q. 1990, \apj, 352, 343

\bibitem[{{MacNeice} {et~al.}(2004){MacNeice}, {Antiochos}, {Phillips},
  {Spicer}, {DeVore}, \& {Olson}}]{MacNeice2004}
{MacNeice}, P., {Antiochos}, S.~K., {Phillips}, A., {et~al.} 2004, \apj, 614,
  1028

\bibitem[{{Moore} {et~al.}(2001){Moore}, {Sterling}, {Hudson}, \&
  {Lemen}}]{Moore2001}
{Moore}, R.~L., {Sterling}, A.~C., {Hudson}, H.~S., \& {Lemen}, J.~R. 2001,
  \apj, 552, 833

\bibitem[{{Myers} {et~al.}(2015){Myers}, {Yamada}, {Ji}, {Yoo}, {Fox},
  {Jara-Almonte}, {Savcheva}, \& {Deluca}}]{Myers2015}
{Myers}, C.~E., {Yamada}, M., {Ji}, H., {et~al.} 2015, \nat, 528, 526

\bibitem[{{Nindos} \& {Andrews}(2004)}]{Nindos2004}
{Nindos}, A., \& {Andrews}, M.~D. 2004, \apjl, 616, L175

\bibitem[{{Nindos} {et~al.}(2012){Nindos}, {Patsourakos}, \&
  {Wiegelmann}}]{Nindos2012}
{Nindos}, A., {Patsourakos}, S., \& {Wiegelmann}, T. 2012, \apjl, 748, L6

\bibitem[{{Olmedo} \& {Zhang}(2010)}]{Olmedo2010}
{Olmedo}, O., \& {Zhang}, J. 2010, \apj, 718, 433

\bibitem[{{Qiu} {et~al.}(2007){Qiu}, {Hu}, {Howard}, \& {Yurchyshyn}}]{Qiu2007}
{Qiu}, J., {Hu}, Q., {Howard}, T.~A., \& {Yurchyshyn}, V.~B. 2007, \apj, 659,
  758

\bibitem[{{Schou} {et~al.}(2012){Schou}, {Borrero}, {Norton}, {Tomczyk},
  {Elmore}, \& {Card}}]{Schou2012}
{Schou}, J., {Borrero}, J.~M., {Norton}, A.~A., {et~al.} 2012, \solphys, 275,
  327

\bibitem[{{Song} {et~al.}(2014){Song}, {Zhang}, {Chen}, \& {Cheng}}]{Song2014}
{Song}, H.~Q., {Zhang}, J., {Chen}, Y., \& {Cheng}, X. 2014, \apjl, 792, L40

\bibitem[{{Sun}(2013)}]{Sun2013}
{Sun}, X. 2013, ArXiv e-prints, arXiv:1309.2392

\bibitem[{{Sun} {et~al.}(2015){Sun}, {Bobra}, {Hoeksema}, {Liu}, {Li}, {Shen},
  {Couvidat}, {Norton}, \& {Fisher}}]{Sun2015}
{Sun}, X., {Bobra}, M.~G., {Hoeksema}, J.~T., {et~al.} 2015, \apjl, 804, L28

\bibitem[{{Toriumi} {et~al.}(2017){Toriumi}, {Schrijver}, {Harra}, {Hudson}, \&
  {Nagashima}}]{Toriumi2017}
{Toriumi}, S., {Schrijver}, C.~J., {Harra}, L.~K., {Hudson}, H., \&
  {Nagashima}, K. 2017, \apj, 834, 56

\bibitem[{{T{\"o}r{\"o}k} \& {Kliem}(2003)}]{Torok2003}
{T{\"o}r{\"o}k}, T., \& {Kliem}, B. 2003, \aap, 406, 1043

\bibitem[{{T{\"o}r{\"o}k} \& {Kliem}(2005)}]{Torok2005}
---. 2005, \apjl, 630, L97

\bibitem[{{T{\"o}r{\"o}k} \& {Kliem}(2007)}]{Torok2007}
---. 2007, Astronomische Nachrichten, 328, 743

\bibitem[{{T{\"o}r{\"o}k} {et~al.}(2004){T{\"o}r{\"o}k}, {Kliem}, \&
  {Titov}}]{Torok2004}
{T{\"o}r{\"o}k}, T., {Kliem}, B., \& {Titov}, V.~S. 2004, \aap, 413, L27

\bibitem[{{Tziotziou} {et~al.}(2012){Tziotziou}, {Georgoulis}, \&
  {Raouafi}}]{Tziotziou2012}
{Tziotziou}, K., {Georgoulis}, M.~K., \& {Raouafi}, N.-E. 2012, \apjl, 759, L4

\bibitem[{{van Ballegooijen} \& {Martens}(1989)}]{vanBallegooijen1989}
{van Ballegooijen}, A.~A., \& {Martens}, P.~C.~H. 1989, \apj, 343, 971

\bibitem[{{Vourlidas} {et~al.}(2010){Vourlidas}, {Howard}, {Esfandiari},
  {Patsourakos}, {Yashiro}, \& {Michalek}}]{Vourlidas2010}
{Vourlidas}, A., {Howard}, R.~A., {Esfandiari}, E., {et~al.} 2010, \apj, 722,
  1522

\bibitem[{{Wang} {et~al.}(2015){Wang}, {Cao}, {Liu}, {Xu}, {Liu}, {Zeng},
  {Chae}, \& {Ji}}]{Wang2015}
{Wang}, H., {Cao}, W., {Liu}, C., {et~al.} 2015, Nature Communications, 6, 7008

\bibitem[{{Wheatland} {et~al.}(2000){Wheatland}, {Sturrock}, \&
  {Roumeliotis}}]{Wheatland2000}
{Wheatland}, M.~S., {Sturrock}, P.~A., \& {Roumeliotis}, G. 2000, \apj, 540,
  1150

\bibitem[{{Wiegelmann}(2004)}]{Wiegelmann2004}
{Wiegelmann}, T. 2004, \solphys, 219, 87

\bibitem[{{Wiegelmann} \& {Inhester}(2010)}]{Wiegelmann2010}
{Wiegelmann}, T., \& {Inhester}, B. 2010, \aap, 516, A107

\bibitem[{{Wiegelmann} {et~al.}(2006){Wiegelmann}, {Inhester}, \&
  {Sakurai}}]{Wiegelmann2006}
{Wiegelmann}, T., {Inhester}, B., \& {Sakurai}, T. 2006, \solphys, 233, 215

\bibitem[{{Wiegelmann} {et~al.}(2012){Wiegelmann}, {Thalmann}, {Inhester},
  {Tadesse}, {Sun}, \& {Hoeksema}}]{Wiegelmann2012}
{Wiegelmann}, T., {Thalmann}, J.~K., {Inhester}, B., {et~al.} 2012, \solphys,
  281, 37

\bibitem[{{Wiegelmann} {et~al.}(2010){Wiegelmann}, {Yelles Chaouche},
  {Solanki}, \& {Lagg}}]{Wiegelmann2010a}
{Wiegelmann}, T., {Yelles Chaouche}, L., {Solanki}, S.~K., \& {Lagg}, A. 2010,
  \aap, 511, A4

\bibitem[{{Xu} {et~al.}(2012){Xu}, {Liu}, {Jing}, \& {Wang}}]{Xu2012b}
{Xu}, Y., {Liu}, C., {Jing}, J., \& {Wang}, H. 2012, \apj, 761, 52

\bibitem[{{Yan} {et~al.}(2015){Yan}, {Xue}, {Pan}, {Wang}, {Xiang}, {Kong}, \&
  {Yang}}]{Yan2015}
{Yan}, X.~L., {Xue}, Z.~K., {Pan}, G.~M., {et~al.} 2015, \apjs, 219, 17

\bibitem[{{Zuccarello} {et~al.}(2015){Zuccarello}, {Aulanier}, \&
  {Gilchrist}}]{Zuccarello2015}
{Zuccarello}, F.~P., {Aulanier}, G., \& {Gilchrist}, S.~A. 2015, \apj, 814, 126

\end{thebibliography}

\newpage

\begin{deluxetable}{c|cccc|c|c|cc|c|c|}
\rotate
\tabletypesize{\scriptsize} \tablecolumns{11} \tablewidth{0pt}
\tablecaption{Event List\label{TAB01}} \tablehead{
    \colhead{No.} &
    \colhead{Flare} &
    \colhead{NOAA} &
    \colhead{Position} &
    \colhead{Type$\tablenotemark{a}$} &
    \colhead{$E_\mathrm{CME}$$\tablenotemark{b}$ } &
    \colhead{$E_\mathrm{free}$$\tablenotemark{c}$ } &
    \colhead{$h_\mathrm{apex}$$\tablenotemark{d}$ } &
    \colhead{$d$$\tablenotemark{e}$ } &
    \colhead{$n$$\tablenotemark{f}$} &
    \colhead{$|T_w|$$\tablenotemark{g}$} \\
    \colhead{} &
    \colhead{SXR peak time, class} &
    \colhead{AR} &
    \colhead{} &
    \colhead{E$/$C} &
    \colhead{($10^{32}$ erg) } &
    \colhead{($10^{32}$ erg) } &
    \colhead{(Mm)} &
    \colhead{(Mm)} &
    \colhead{} &
    \colhead{}}
\startdata
1  & SOL2011-02-13T17:38, M6.6 & 11158 & S20E04 & E  & 0.007 & 1.1$\pm$0.4   &  5.4$\pm$0.7  & 18.6$\pm$2.0  & 0.58$\pm$0.16 & 0.39$\pm$0.03\\
2  & SOL2011-02-15T01:56, X2.2 & 11158 & S20W10 & E  & 0.1   & 2.6$\pm$1.0   & 13.1$\pm$1.8  & 28.4$\pm$2.0  & 0.86$\pm$0.46 & 0.62$\pm$0.04\\
3  & SOL2011-03-09T23:23, X1.5 & 11166 & N08W09 & C  & $--$  & 2.2$\pm$0.9   & 12.8$\pm$1.3  & 33.9$\pm$1.4  & 0.62$\pm$0.04 & 0.73$\pm$0.21\\
4  & SOL2011-07-30T02:09, M9.3 & 11261 & S20W10 & C  & $--$  & 1.1$\pm$0.4   &  7.1$\pm$0.6  & 20.0$\pm$0.6  & 0.45$\pm$0.12 & 0.47$\pm$0.14\\
5  & SOL2011-08-03T13:48, M6.0 & 11261 & N16W30 & E  & 0.2   & 2.1$\pm$0.8   & 18.1$\pm$1.9  & 54.8$\pm$1.3  & 1.28$\pm$0.12 & 0.73$\pm$0.16\\
6  & SOL2011-09-06T01:50, M5.3 & 11283 & N14W07 & E  & 0.07  & 0.8$\pm$0.3   & 10.7$\pm$0.9  & 35.7$\pm$2.5  & 0.86$\pm$0.14 & 0.63$\pm$0.09\\
7  & SOL2011-09-06T22:20, X2.1 & 11283 & N14W18 & E  & 0.3   & 1.0$\pm$0.4   & 22.2$\pm$0.9  & 35.9$\pm$7.4  & 0.98$\pm$0.33 & 0.98$\pm$0.20\\
8  & SOL2011-10-02T00:50, M3.9 & 11305 & N12W26 & C  & $--$  & 1.0$\pm$0.4   &  7.6$\pm$1.1  & 21.5$\pm$5.0  & 0.63$\pm$0.08 & 0.55$\pm$0.09\\
9  & SOL2012-01-23T03:59, M8.7 & 11402 & N28W21 & E  & 6.2   & 3.0$\pm$1.2   & 19.4$\pm$0.7  & 31.1$\pm$0.9  & 0.79$\pm$0.11 & 1.08$\pm$0.20\\
10 & SOL2012-03-07T00:24, X5.4 & 11429 & N17E31 & E  & 5.0   & 9.3$\pm$3.6   & 14.4$\pm$0.9  & 30.0$\pm$2.0  & 0.89$\pm$0.11 & 0.62$\pm$0.06\\

11 & SOL2012-03-09T03:53, M6.3 & 11429 & N15W03 & E  & 0.3   & 6.8$\pm$2.7   & 10.6$\pm$0.6  & 65.6$\pm$8.7  & 0.78$\pm$0.09 & 0.78$\pm$0.15\\
12 & SOL2012-07-02T10:52, M5.6 & 11515 & S17E08 & E  & 0.02  & 1.4$\pm$0.5   & 12.6$\pm$0.7  & 33.0$\pm$3.7  & 0.60$\pm$0.18 & 1.44$\pm$0.16\\
13 & SOL2012-07-12T16:49, X1.4 & 11520 & S15W01 & E  & 0.3   & 5.8$\pm$2.3   &  9.3$\pm$0.7  & 32.8$\pm$0.8  & 0.51$\pm$0.06 & 0.81$\pm$0.09\\
14 & SOL2013-04-11T07:16, M6.5 & 11719 & N09E12 & E  & 0.8   & 0.5$\pm$0.2   & 13.5$\pm$1.6  & 37.0$\pm$3.1  & 0.61$\pm$0.12 & 0.92$\pm$0.17\\
15 & SOL2013-10-24T00:30, M9.3 & 11877 & S09E10 & E  & 0.02  & 1.5$\pm$0.6   &  8.3$\pm$0.4  & 30.2$\pm$7.4  & 0.18$\pm$0.20 & 0.99$\pm$0.12\\
16 & SOL2013-11-01T19:53, M6.3 & 11884 & S12E01 & C  & $--$  & 1.2$\pm$0.5   &  8.3$\pm$0.3  & 34.6$\pm$1.7  & 0.70$\pm$0.23 & 1.15$\pm$0.21\\
17 & SOL2013-11-03T05:22, M4.9 & 11884 & S12W17 & C  & $--$  & 0.9$\pm$0.4   &  8.8$\pm$2.8  & 19.2$\pm$7.2  & 0.53$\pm$0.12 & 0.81$\pm$0.14\\
18 & SOL2013-11-05T22:12, X3.3 & 11890 & S12E44 & E  & 0.08  & 1.9$\pm$0.7   &  7.7$\pm$0.4  & 29.0$\pm$0.5  & 0.31$\pm$0.18 & 0.54$\pm$0.14\\
19 & SOL2013-11-08T04:26, X1.1 & 11890 & S12E13 & E  & 0.1   & 1.1$\pm$0.4   &  8.3$\pm$0.4  & 20.1$\pm$2.9  & 0.28$\pm$0.19 & 0.59$\pm$0.08\\
20 & SOL2014-01-07T18:32, X1.2 & 11944 & S15W11 & E  & 3.7   & 6.0$\pm$2.3   & 19.9$\pm$0.8  & 97.4$\pm$1.1  & 0.61$\pm$0.07 & 0.84$\pm$0.15\\

21 & SOL2014-02-02T09:31, M4.4 & 11967 & S10E13 & C  & $--$  & 6.7$\pm$2.6   &  7.8$\pm$0.9  & 31.0$\pm$1.3  & 0.43$\pm$0.05 & 0.66$\pm$0.09\\
22 & SOL2014-02-04T04:00, M5.2 & 11967 & S14W06 & C  & $--$  & 6.3$\pm$2.5   &  8.0$\pm$1.1  & 20.5$\pm$5.4  & 0.67$\pm$0.09 & 0.67$\pm$0.11\\
23 & SOL2014-03-29T17:48, X1.1 & 12017 & N10W32 & E  & 0.07  & 1.5$\pm$0.6   &  7.9$\pm$0.4  & 36.9$\pm$3.4  & 0.73$\pm$0.19 & 1.26$\pm$0.09\\
24 & SOL2014-04-18T13:03, M7.3 & 12036 & S20W34 & E  & 1.4   & 3.1$\pm$1.2   & 25.8$\pm$2.7  & 38.2$\pm$3.3  & 1.07$\pm$0.10 & 0.75$\pm$0.12\\
25 & SOL2014-09-10T17:45, X1.6 & 12158 & N11E05 & E  & 1.7   & 2.6$\pm$1.0   &  8.2$\pm$1.2  & 46.2$\pm$2.5  & 0.42$\pm$0.06 & 0.49$\pm$0.09\\
26 & SOL2014-10-22T14:28, X1.6 & 12192 & S14E13 & C  & $--$  & 22.8$\pm$8.9  &  5.5$\pm$0.4  & 35.0$\pm$1.0  & 0.22$\pm$0.04 & 0.72$\pm$0.03\\
27 & SOL2014-10-24T21:41, X3.1 & 12192 & S22W21 & C  & $--$  & 29.2$\pm$11.4 & 23.8$\pm$2.9  & 101.6$\pm$12.7& 0.70$\pm$0.08 & 0.97$\pm$0.26\\
28 & SOL2014-11-07T17:26, X1.6 & 12205 & N17E40 & E  & 0.4   & 4.2$\pm$1.6   &  7.7$\pm$0.9  & 44.7$\pm$0.8  & 0.66$\pm$0.13 & 1.03$\pm$0.11\\
29 & SOL2014-12-04T18:25, M6.1 & 12222 & S20W31 & C  & $--$  & 0.7$\pm$0.3   & 12.9$\pm$0.3  & 55.4$\pm$1.1  & 0.56$\pm$0.10 & 1.05$\pm$0.17\\
30 & SOL2014-12-17T04:51, M8.7 & 12242 & S18E08 & E  & 0.2   & 5.9$\pm$2.3   & 27.5$\pm$1.8  & 64.6$\pm$0.6  & 1.34$\pm$0.10 & 0.71$\pm$0.12\\

31 & SOL2014-12-18T21:58, M6.9 & 12241 & S11E15 & E  & N/A\tablenotemark{h} & 2.2$\pm$0.9    & 43.6$\pm$0.4  & 36.9$\pm$3.4  & 1.57$\pm$0.10 & 1.08$\pm$0.14\\
32 & SOL2014-12-20T00:28, X1.8 & 12242 & S19W29 & E  & 0.8   & 11.8$\pm$4.6  & 10.6$\pm$0.4  & 41.7$\pm$1.2  & 0.65$\pm$0.06 & 0.91$\pm$0.09\\
33 & SOL2015-03-12T14:08, M4.2 & 12297 & S15E06 & C  & $--$  & 2.9$\pm$1.1   &  3.2$\pm$0.3  & 25.0$\pm$3.8  & 0.36$\pm$0.15 & 0.54$\pm$0.08\\
34 & SOL2015-06-22T18:23, M6.5 & 12371 & N13W06 & E  & 0.3   & 6.9$\pm$2.7   & 12.2$\pm$1.8  & 44.2$\pm$2.3  & 0.92$\pm$0.06 & 1.14$\pm$0.05\\
35 & SOL2015-06-25T08:16, M7.9 & 12371 & N12W40 & E  & 4.1   & 6.9$\pm$2.7   & 20.1$\pm$0.6  & 42.2$\pm$2.6  & 0.74$\pm$0.20 & 0.93$\pm$0.07\\
36 & SOL2015-09-28T14:58, M7.6 & 12422 & S20W28 & C  & $--$  & 1.7$\pm$0.7   & 10.3$\pm$0.2  & 42.7$\pm$7.3  & 0.44$\pm$0.13 & 0.74$\pm$0.21\\
37 & SOL2017-09-04T20:33, M5.5 & 12673 & S10W11 & E  & N/A\tablenotemark{i} & 5.4$\pm$2.1    &  2.4$\pm$0.5  & 16.9$\pm$1.5  & 0.80$\pm$0.27 & 0.70$\pm$0.08\\
38 & SOL2017-09-06T12:02, X9.3 & 12673 & S09W34 & E  & N/A\tablenotemark{i} & 1.3$\pm$0.5    &  5.0$\pm$0.5  & 18.5$\pm$4.5  & 0.6w3$\pm$0.09 & 0.79$\pm$0.14\\

\enddata

\tablenotetext{a}{Flare type: (E)jective or (C)onfined;}
\tablenotetext{b}{CME Kinetic energy, from \emph{LASCO} CME catalog;}
\tablenotetext{c}{Free magnetic energy with a 39\% uncertainty (see \S 2); }
\tablenotetext{d}{Apex height of MFRs with $\pm1\sigma$ uncertainty obtained from NLFF field lines forming the MFRs;}
\tablenotetext{e}{Footpoint distance of MFRs with $\pm1\sigma$ uncertainty obtained from NLFF field lines forming the MFRs;}
\tablenotetext{f}{Decay index with $\pm1\sigma$ uncertainty calculated over the FPIL mask (see \S 3.1);}
\tablenotetext{g}{Twist number with $\pm1\sigma$ uncertainty calculated over the NLFF field lines forming the MFRs (see \S 3.2);}
\tablenotetext{h}{Not available due to the data gap of \emph{LASCO} C2;}
\tablenotetext{i}{Not available in the \emph{LASCO} CME catalog.}
\end{deluxetable}

\begin{figure}
\begin{center}
\begin{tabular}{cc}
\resizebox{75mm}{!}{\includegraphics{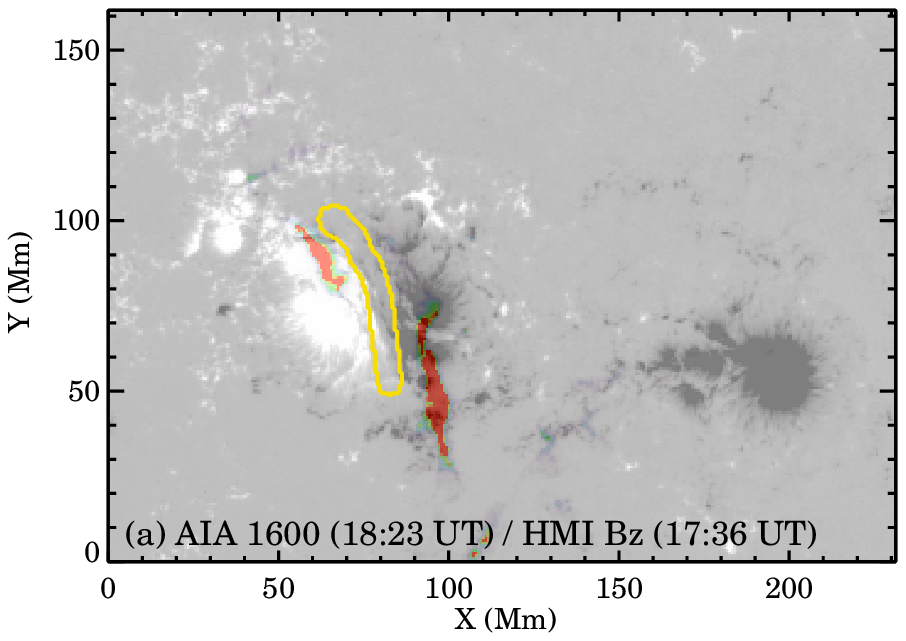}}&
\resizebox{75mm}{!}{\includegraphics{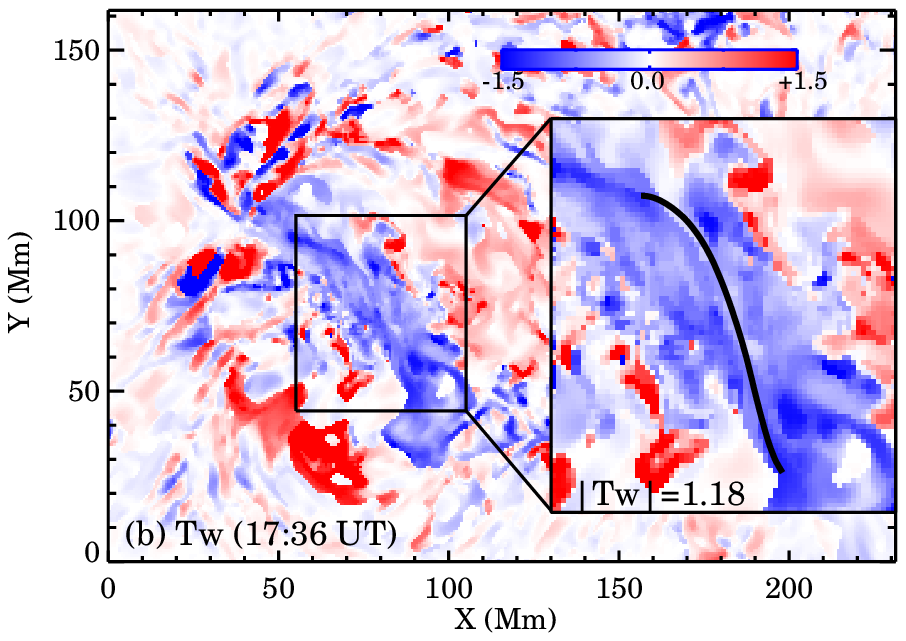}}\\
\resizebox{75mm}{!}{\includegraphics{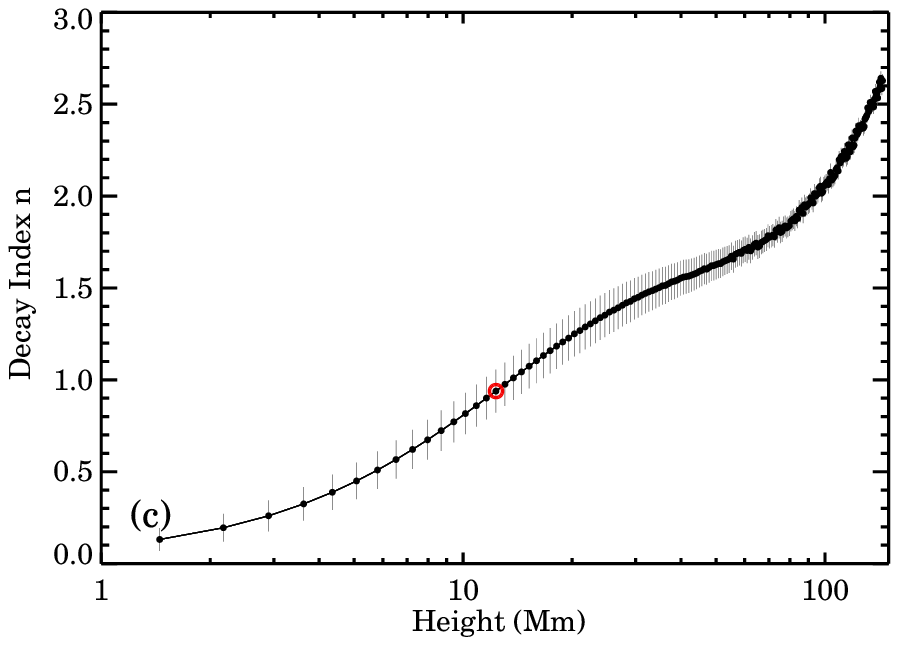}}&
\resizebox{75mm}{!}{\includegraphics{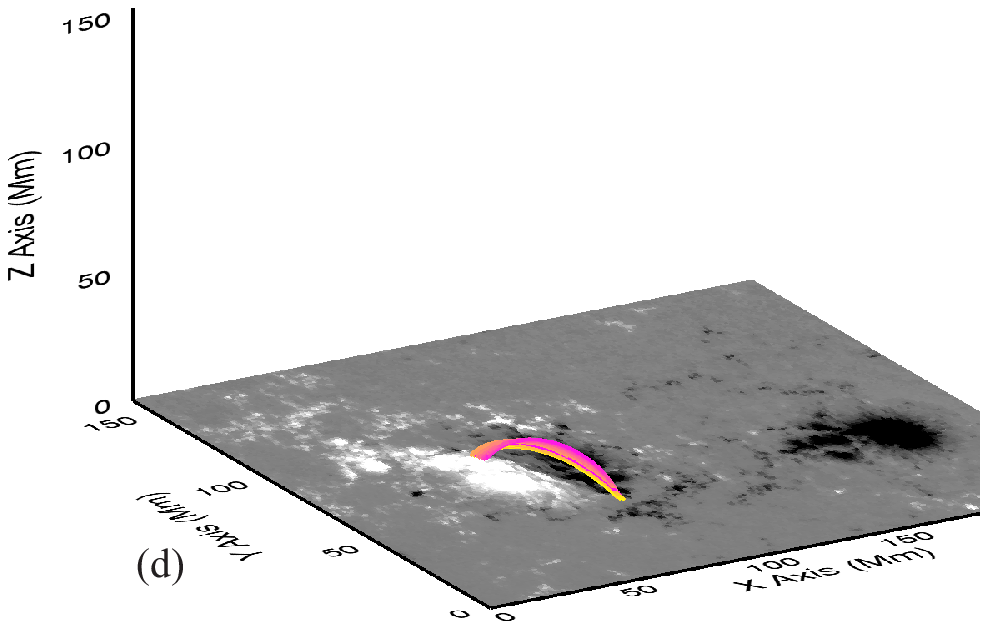}} \\
\end{tabular}
\caption{The magnetic field of the eruptive M6.5 flare (SOL2015-06-22T18:23) of AR 12371. (a): A blend of an AIA UV 1600\AA\ image at the flare peak time with the pre-flare HMI vector magnetogram $B_z$, superimposed with the yellow contours of the flaring polarity inversion line (FPIL) mask. Both AIA and HMI maps are de-rotated to a reference pre-flare time (17:36 UT in this case) and re-mapped with the CEA projection. (b): The twist number $T_w$ map derived from the NLFF field, scaled between $-/+$1.5 (blue/red). The rectangle enclosing the flaring core region is zoomed in and displayed in the inset. The superimposed black line shows a representative field line of the MFR, whose $|T_w|$ is annotated. (c) The height profile of decay index $n$ above the FPIL region derived from the potential field model. The error bars indicate $\pm1\sigma$ spread, evaluated from 908 profiles in FPIL region in this case. The red circle marks the data point at $h_{apex}$. (d) a 3D perspective of the MFR extrapolated from the NLFF field.} \label{fig01}
\end{center}
\end{figure}

\begin{figure}
\begin{center}
\begin{tabular}{cc}
\resizebox{75mm}{!}{\includegraphics{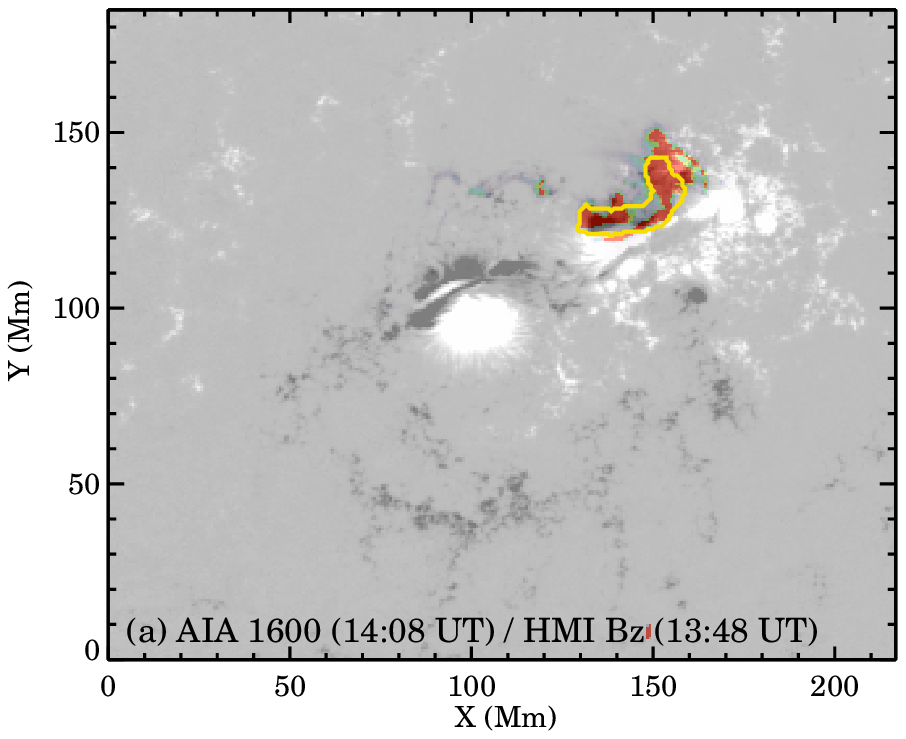}} &
\resizebox{75mm}{!}{\includegraphics{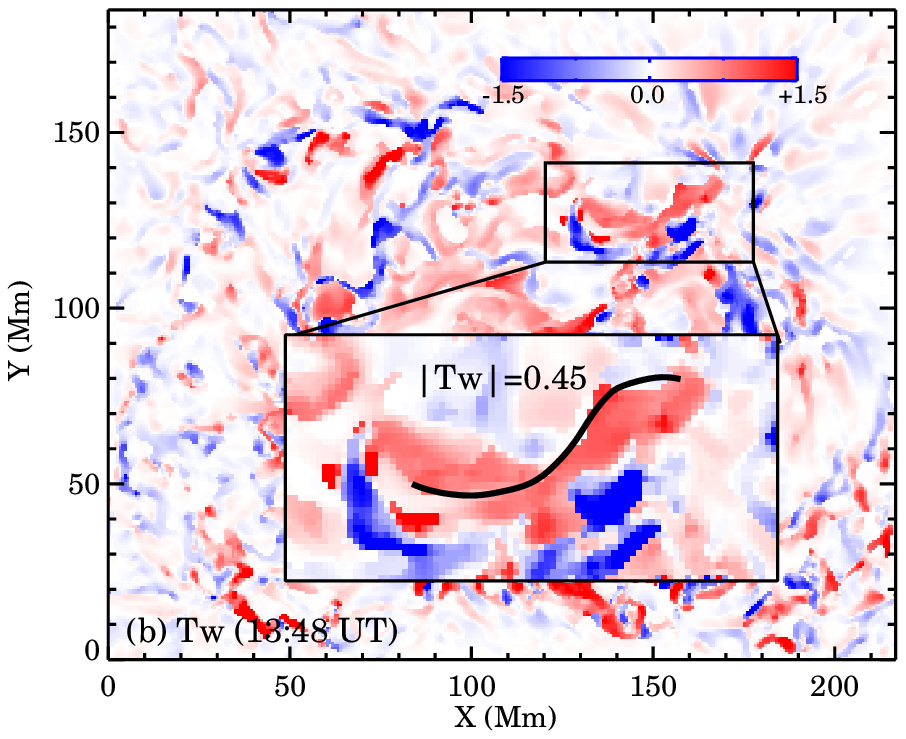}}\\
\resizebox{75mm}{!}{\includegraphics{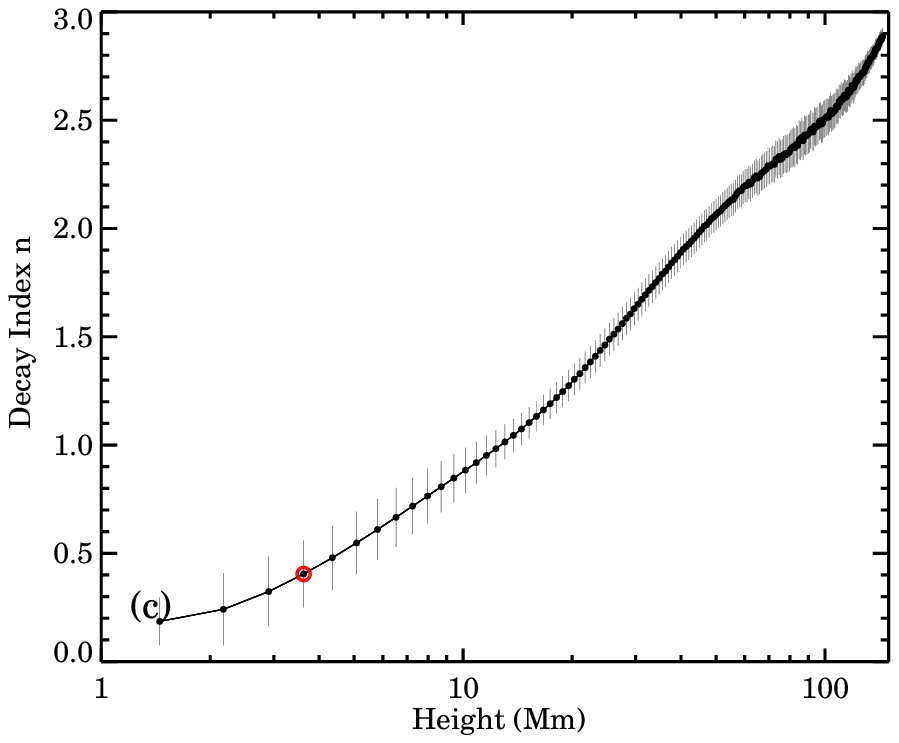}} &
\resizebox{75mm}{!}{\includegraphics{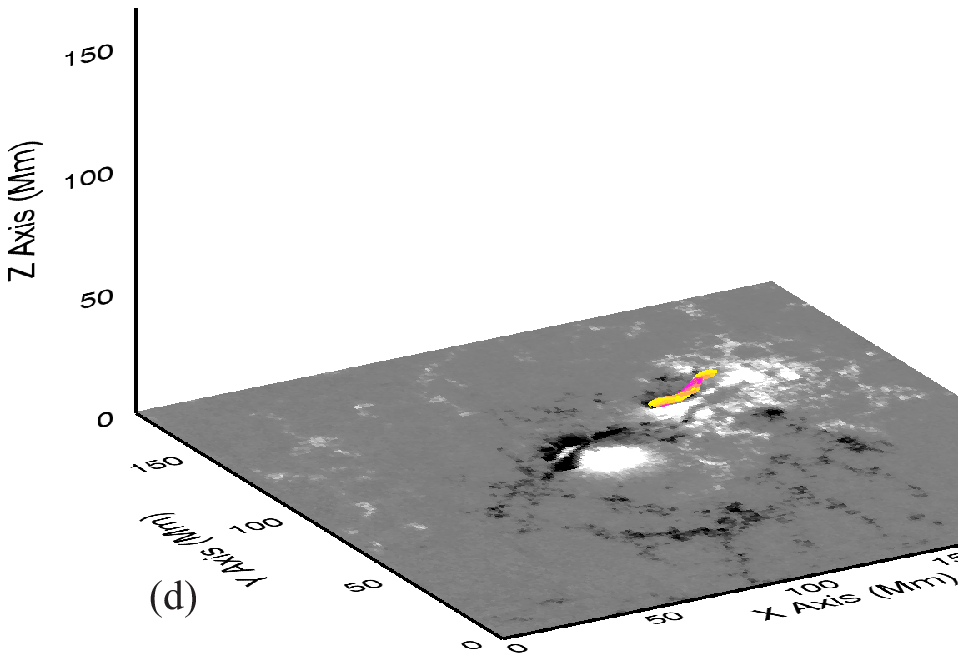}} \\
\end{tabular}
\caption{The magnetic field of the confined M4.2 flare (SOL2015-03-12T14:08) of AR 12297. Same layout as Figure 1.} \label{fig02}
\end{center}
\end{figure}

\begin{figure}
\begin{center}
\epsscale{1.} \plotone{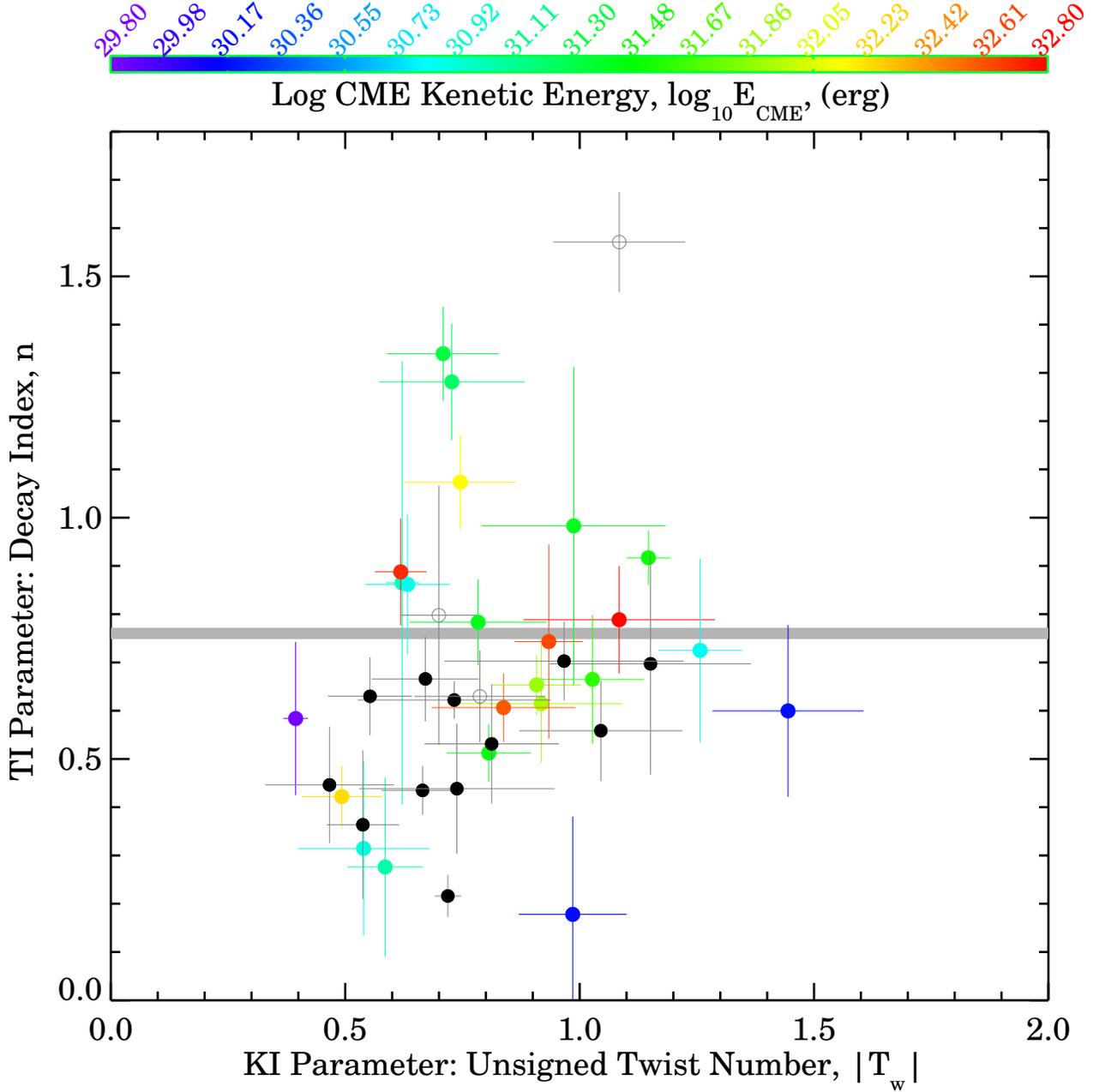}
\caption{Scatter diagram of TI prarameter $n$ vs. KI parameter $|T_w|$. Black and colored symbols correspond to the confined and ejective flares, respectively. For ejective flares, the color is assigned according to the associated CME's kinetic energy, indicated by the color code. Three uncolored hollow symbols represent the three ejective flares in the absence of $E_\mathrm{CME}$ information. The error bars indicate $\pm1\sigma$ spread. The horizontal grey line is drawn to illustrate $n_\mathrm{crit}\simeq0.8$.} \label{fig03}
\end{center}
\end{figure}

\begin{figure}
\begin{center}
\begin{tabular}{ccc}
\resizebox{50mm}{!}{\includegraphics{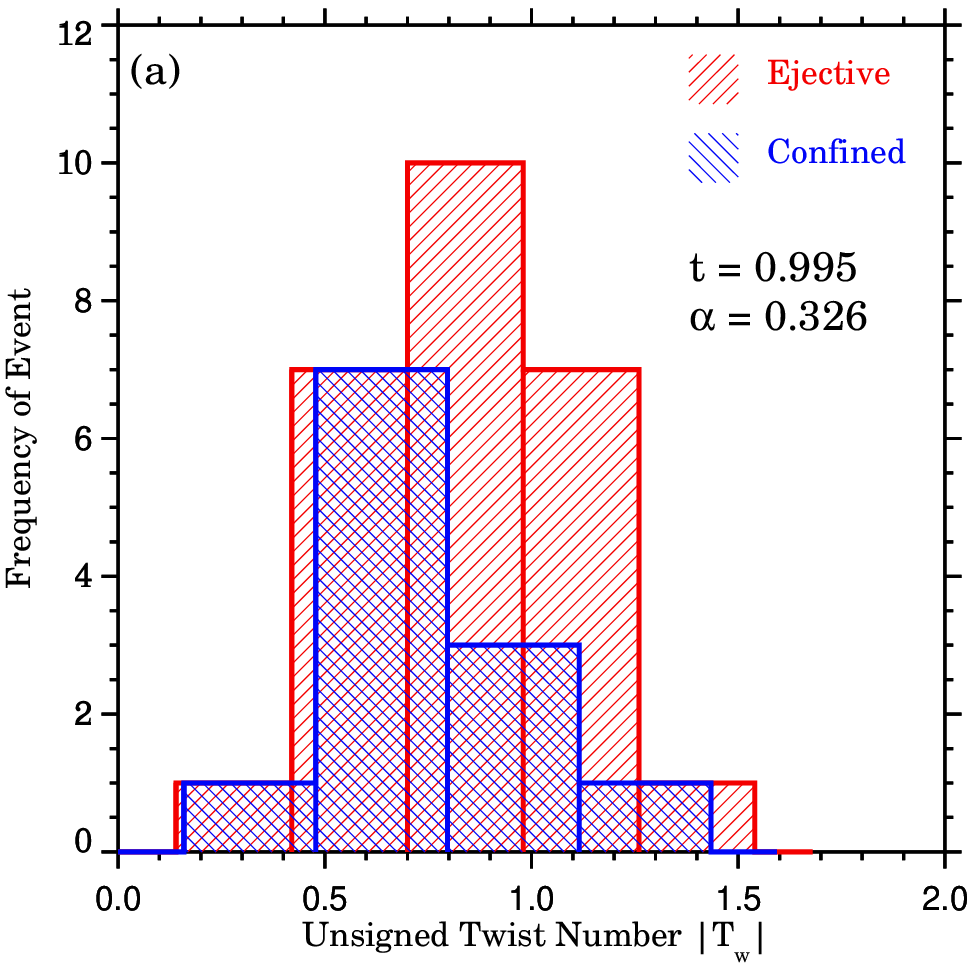}}&
\resizebox{50mm}{!}{\includegraphics{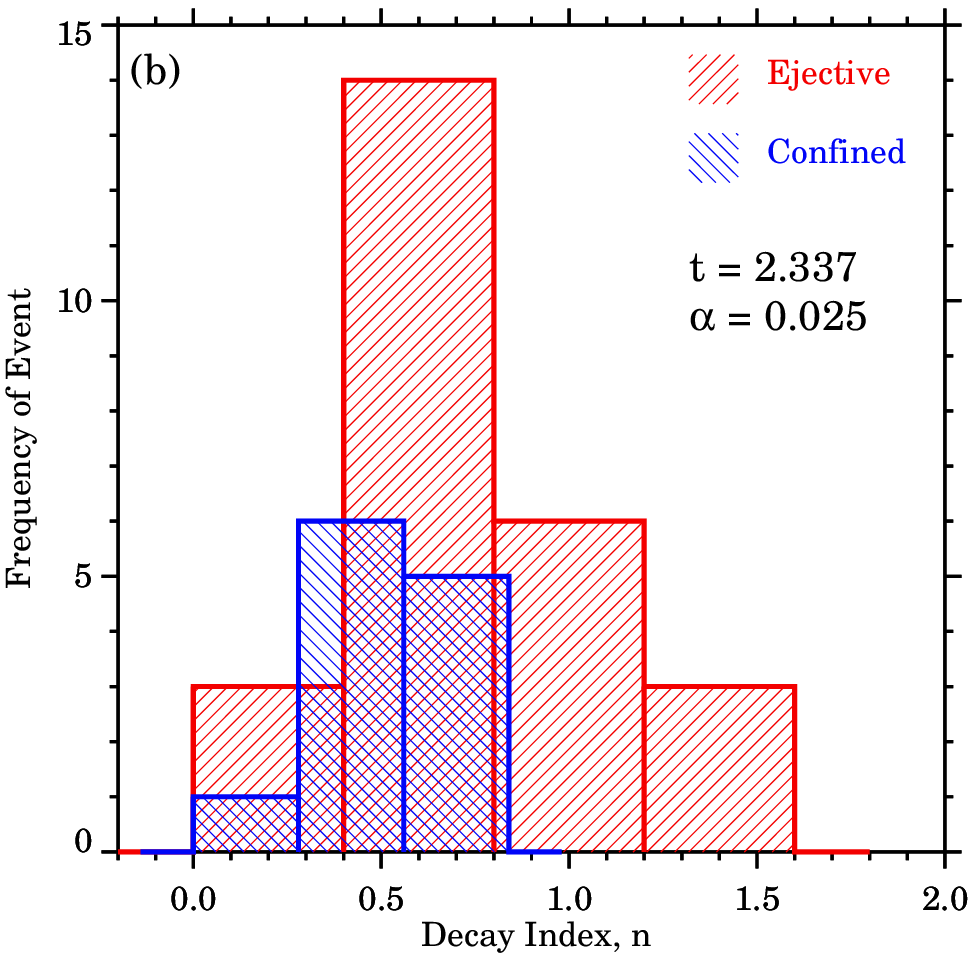}}&
\resizebox{50mm}{!}{\includegraphics{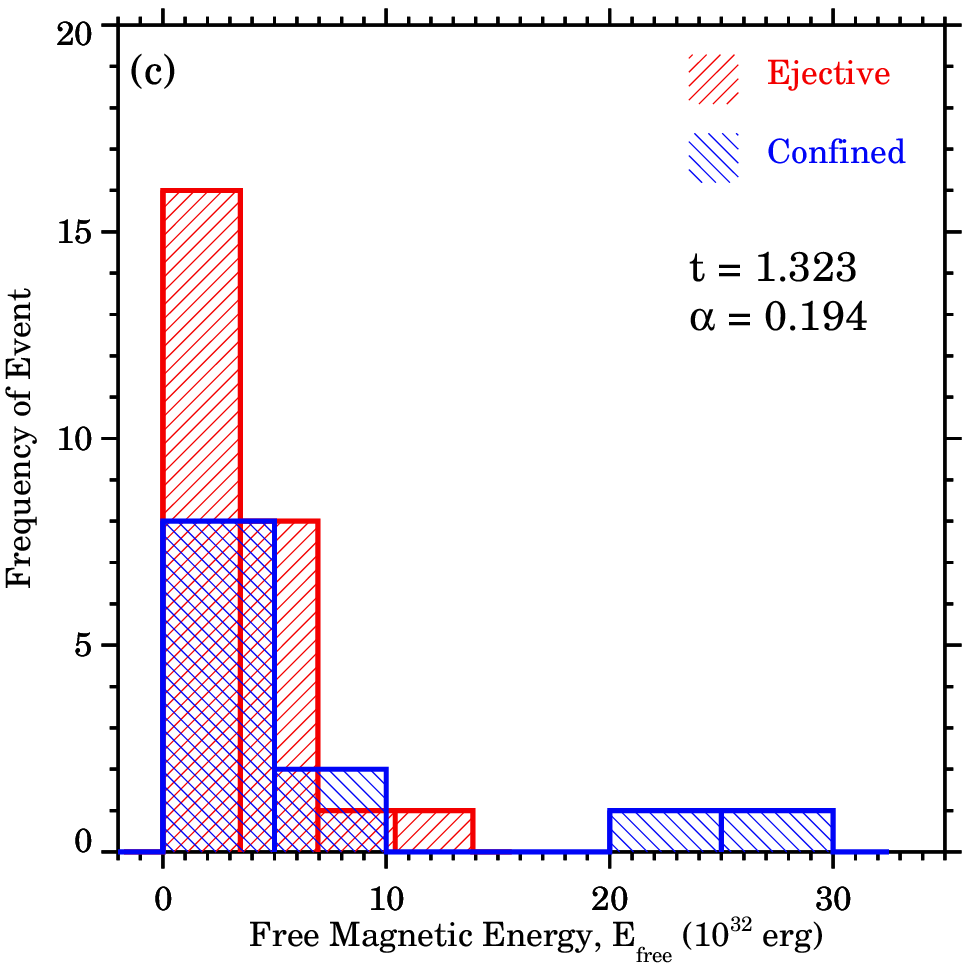}}\\
\resizebox{50mm}{!}{\includegraphics{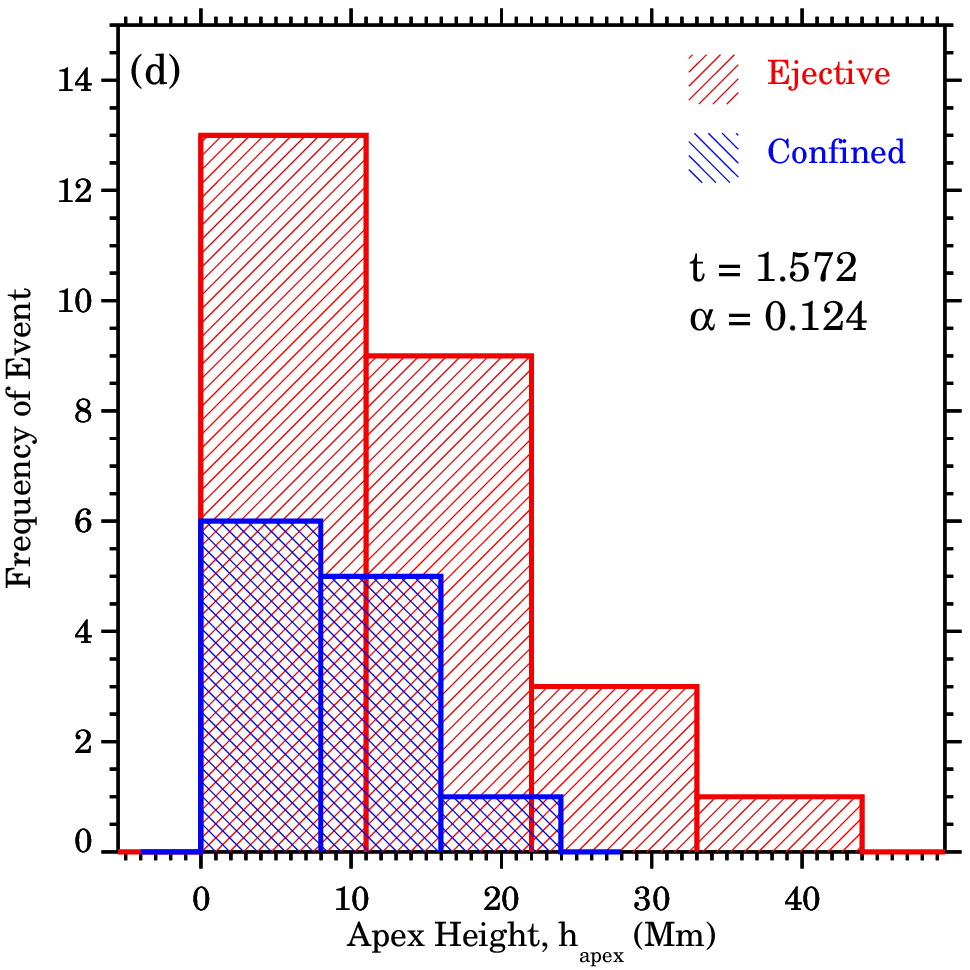}}&
\resizebox{50mm}{!}{\includegraphics{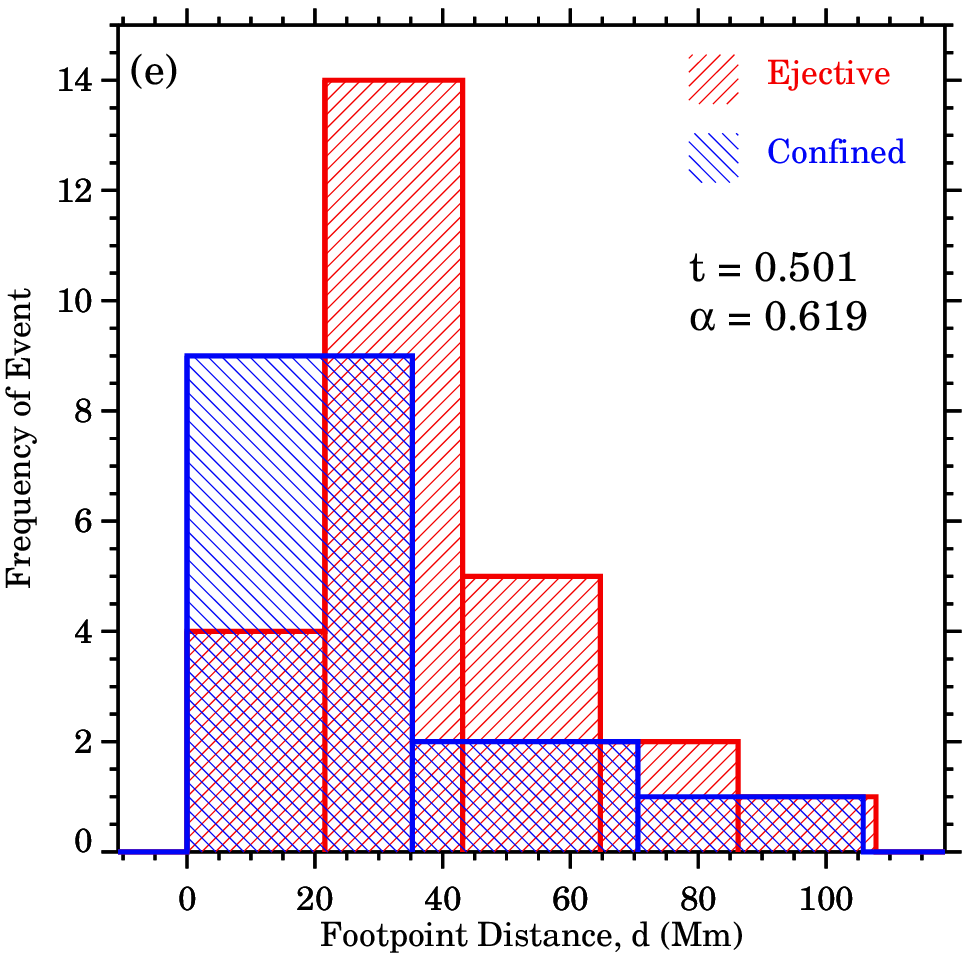}}&
\resizebox{50mm}{!}{\includegraphics{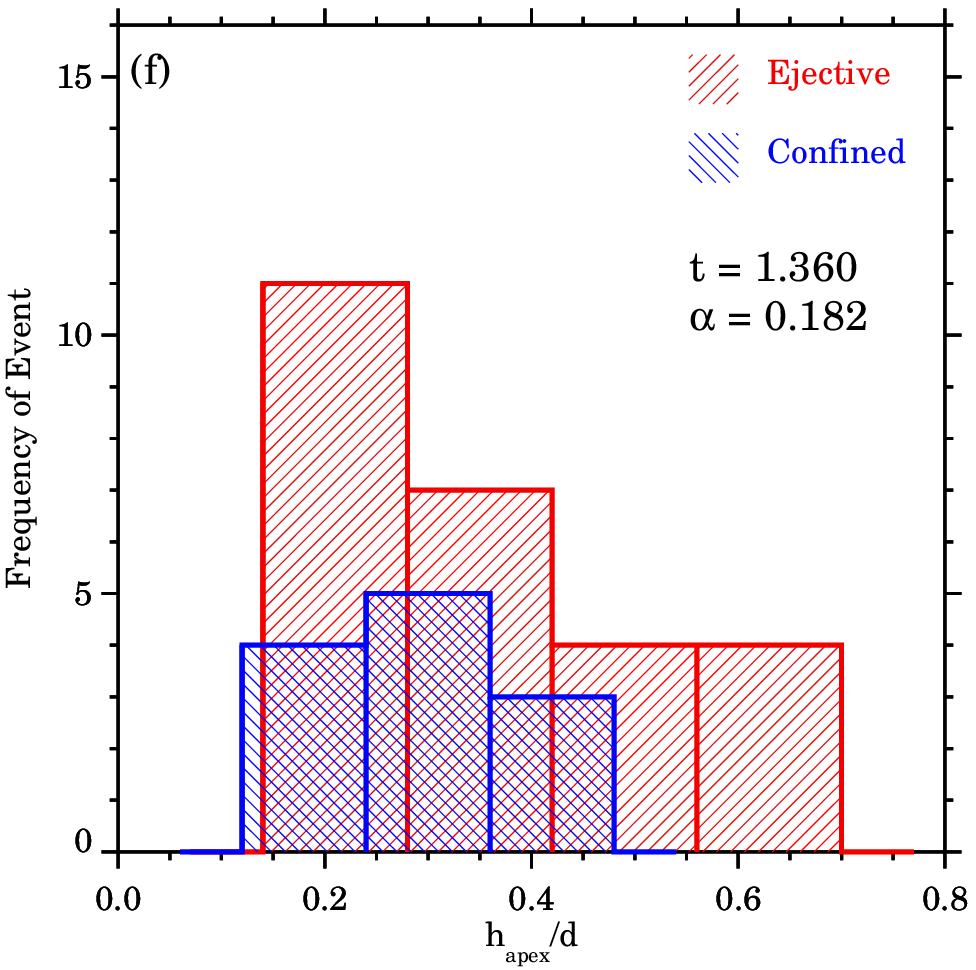}}\\
\end{tabular}
\caption{Histograms of (a) $|T_w|$, (b) $n$, (c) $E_\mathrm{free}$, (d) $h_\mathrm{apex}$, (e)d, and (f)$h_\mathrm{apex}/d$. Red/Blue represents ejective/confined flares.  Student's t-statistic ($t$) and its significance ($\alpha$) are shown in each panel.} \label{fig04}
\end{center}
\end{figure}

\begin{figure}
\begin{center}
\begin{tabular}{cc}
\resizebox{60mm}{!}{\includegraphics{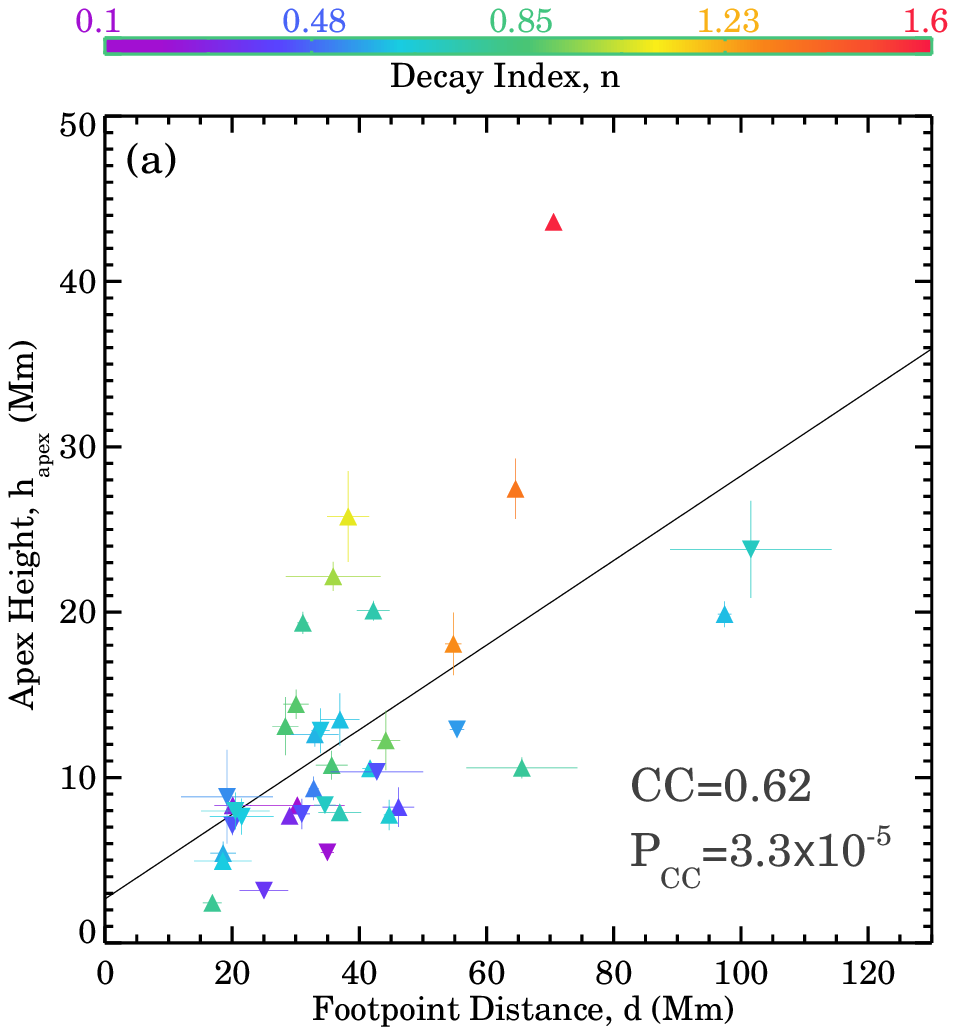}}&
\resizebox{60mm}{!}{\includegraphics{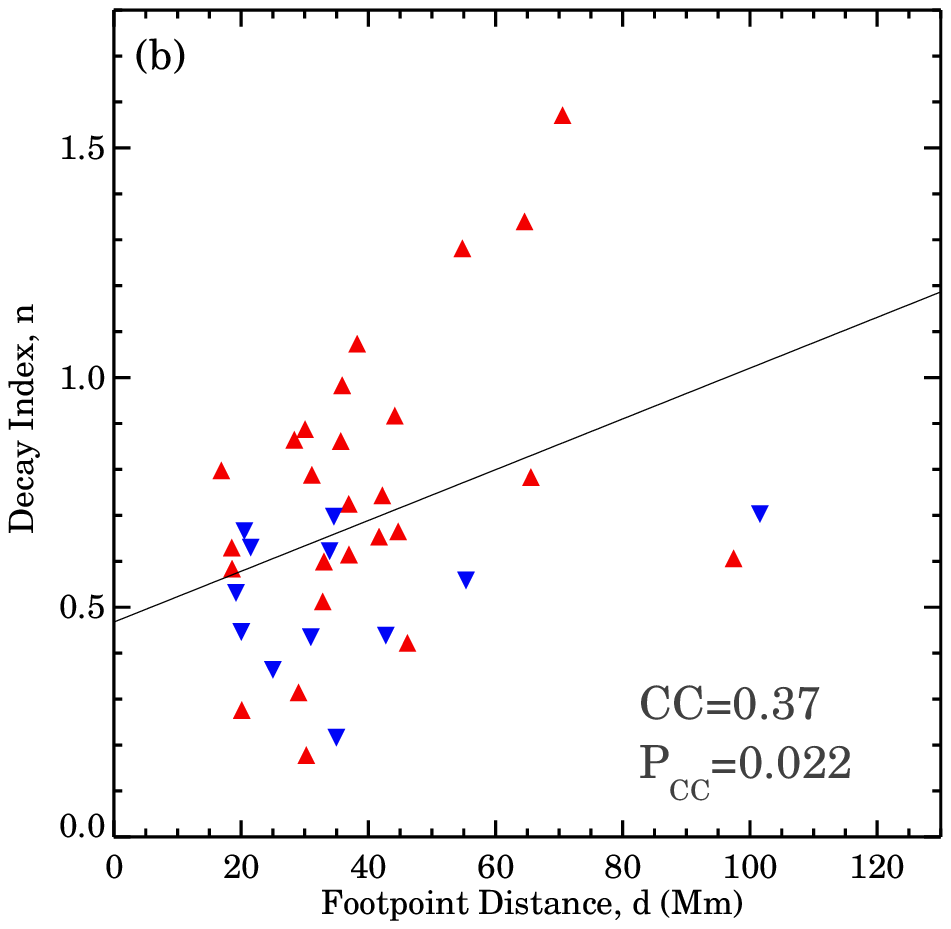}}\\
\resizebox{60mm}{!}{\includegraphics{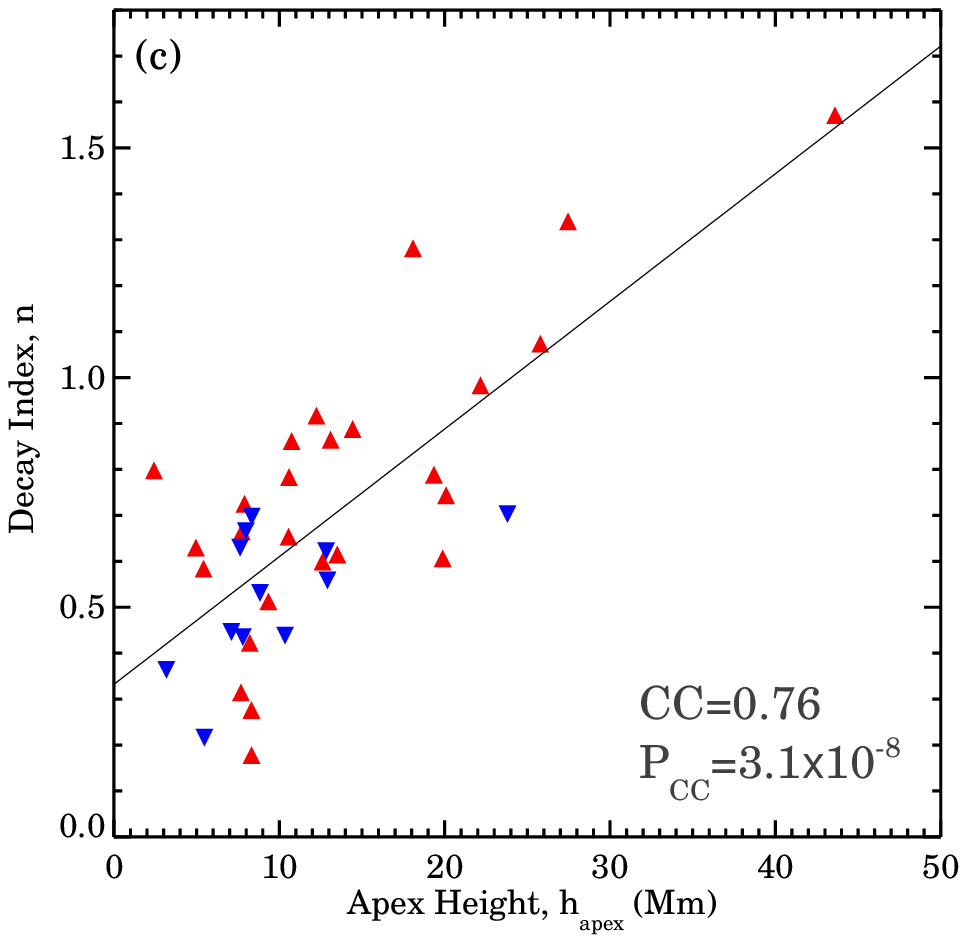}}&
\resizebox{60mm}{!}{\includegraphics{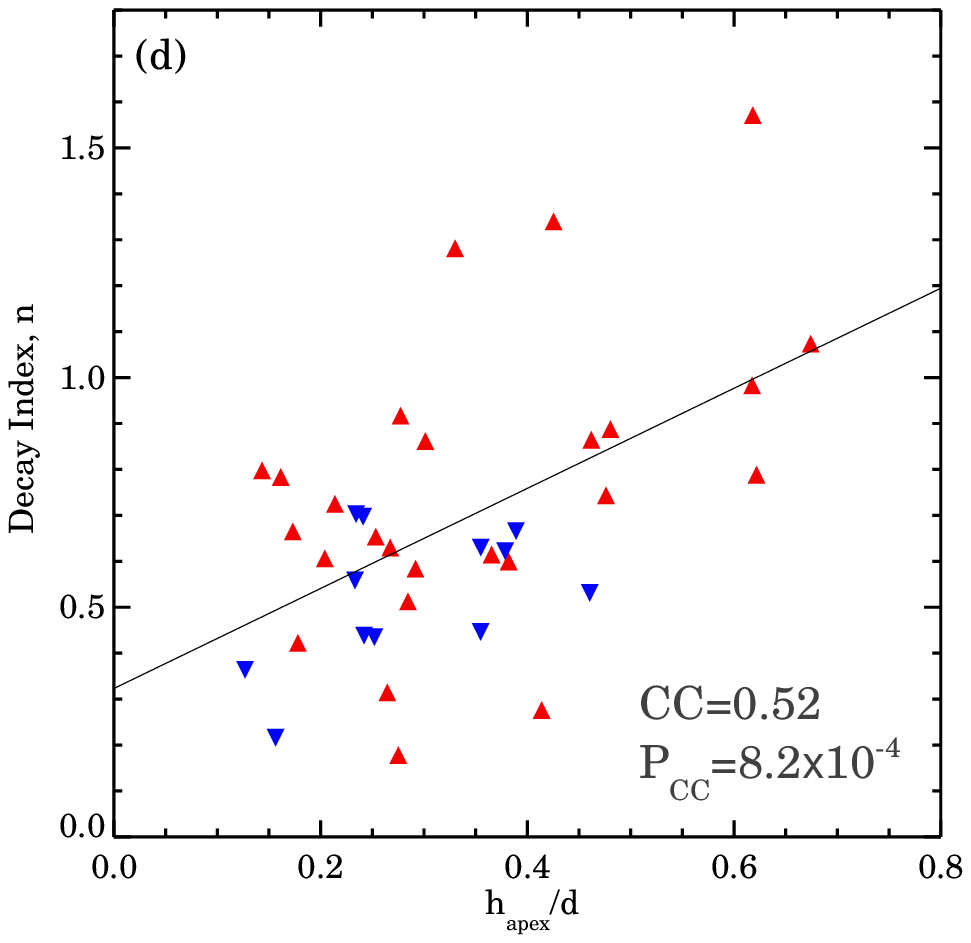}}\\
\end{tabular}
\caption{Scatter diagrams of (a) apex height $h_\mathrm{apex}$ vs. footpoint distance $d$ of MFRs, (b) $n$ vs. $d$, (c) $n$ vs. $h_\mathrm{apex}$, and (d) $n$ vs. $h_\mathrm{apex}/d$. Triangles and up-side-down triangles represent ejective and confined flares, respectively. The color is assigned either according to the value of $n$ (panel a) indicated by the color code, or red/blue for ejective/confined flares (panels b-c). The linear Pearson correlation coefficient (CC) and the probability of obtaining a certain CC by chance ($P_{CC}$) are shown in each panel. The solid lines denote the least-squares fits to data pairs, which are $h_\mathrm{apex}=2.65+0.26\times d$, $n=0.47+0.0055\times d$, $n=0.33+0.028\times h_\mathrm{apex}$, and $n=0.32+1.09\times h_\mathrm{apex}/d$. } \label{fig05}
\end{center}
\end{figure}

\end{document}